\documentclass[preprint2]{aastex}

\slugcomment{To be submitted to the ApJ}

\shorttitle{BDs in NCGC~3603}
\shortauthors{Spezzi et al.}

\begin{document}

\title{Detection of brown dwarf-like objects in the core of NGC~3603}

\author{ 
Loredana Spezzi\altaffilmark{1},
Giacomo Beccari\altaffilmark{1}, 
Guido De Marchi\altaffilmark{1},
Erick T.~Young\altaffilmark{2},
Francesco Paresce\altaffilmark{3},
Michael A.~Dopita\altaffilmark{4},
Morten Andersen\altaffilmark{1},
Nino Panagia\altaffilmark{5},
Bruce Balick\altaffilmark{6},
Howard E.~Bond\altaffilmark{5},
Daniela Calzetti\altaffilmark{7},
C. Marcella Carollo\altaffilmark{8},
Michael J.~Disney\altaffilmark{9},
Jay A.~Frogel\altaffilmark{10},
Donald N.~B.~Hall\altaffilmark{11}, 
Jon A.~Holtzman\altaffilmark{12},
Randy A.~Kimble\altaffilmark{13},
Patrick J.~McCarthy\altaffilmark{14},
Robert W.~O'Connell\altaffilmark{15},
Abhijit Saha\altaffilmark{16},
Joseph I.~Silk\altaffilmark{17},
John T.~Trauger\altaffilmark{18},
Alistair R.~Walker\altaffilmark{19},
Bradley C.~Whitmore\altaffilmark{5},
Rogier A.~Windhorst\altaffilmark{20}
}

\altaffiltext{1}{European Space Agency (ESTEC), PO Box 299, 2200 AG Noordwijk, The Netherlands. e-mail: lspezzi@rssd.esa.int}
\altaffiltext{2}{SOFIA Science Center, NASA Ames Research Center, Moffett Field, California 94035, USA}
\altaffiltext{3}{Istituto di Fisica Spaziale e Fisica Cosmica - Bologna, via Gobetti 101, 40129 Bologna, Italy}
\altaffiltext{4}{Research School of Astronomy \& Astrophysics, ÊThe ÊAustralian National University, ACT 2611, Australia}
\altaffiltext{5}{Space Telescope Science Institute, 3700 San Martin Drive, Baltimore, MD 21218, USA}
\altaffiltext{6}{Department of Astronomy, University of Washington, Seattle, WA 98195-1580, USA}
\altaffiltext{7}{Department of Astronomy, University of Massachusetts, Amherst, MA 01003, USA}
\altaffiltext{8}{Department of Physics, ETH-Zurich, Zurich, 8093, Switzerland}
\altaffiltext{9}{School of Physics and Astronomy, Cardiff University, Cardiff CF24 3AA, United Kingdom}
\altaffiltext{10}{Association of Universities for Research in Astronomy, Washington, DC 20005, USA}
\altaffiltext{11}{Institute for Astronomy, University of Hawaii, Honolulu, HI 96822, USA}
\altaffiltext{12}{Department of Astronomy, New Mexico State University, Las Cruces, NM 88003, USA}
\altaffiltext{13}{NASA--Goddard Space Flight Center, Greenbelt, MD 20771, USA}
\altaffiltext{14}{Observatories of the Carnegie Institution of Washington, Pasadena, CA 91101-1292, USA}
\altaffiltext{15}{Department of Astronomy, University of Virginia, Charlottesville, VA 22904-4325, USA}
\altaffiltext{16}{National Optical Astronomy Observatories, Tucson, AZ 85726-6732, USA}
\altaffiltext{17}{Department of Physics, University of Oxford, Oxford ÊOX1 3PU, United Kingdom}
\altaffiltext{18}{NASA--Jet Propulsion Laboratory, Pasadena, CA 91109, USA}
\altaffiltext{19}{Cerro Tololo Inter-American Observatory, La Serena, Chile}
\altaffiltext{20}{School of Earth and Space Exploration, Arizona State University, Tempe, AZ 85287-1404, USA}

\begin{abstract}

We used near-infrared data obtained with the Wide Field Camera~3 (WFC3) on the 
\emph{Hubble Space Telescope} to identify objects having the colors of brown dwarfs (BDs) in the field of the 
massive galactic cluster NGC~3603. These are identified through a combination of narrow 
and medium band filters which span the J and H bands and are particularly sensitive to 
the presence of the 1.3-1.5$\mu$m H$_2$O molecular band unique to BDs. 
We provide a calibration of the relationship between effective temperature and color for both field 
stars and for BDs. This photometric method provides effective temperatures for BDs to an accuracy of $\pm$350K relative to 
spectroscopic techniques. This accuracy is shown to be not significantly affected by either stellar 
surface gravity or uncertainties in the interstellar extinction. We identify nine objects having 
effective temperature between 1700 and 2200~K, typical of BDs, observed J-band magnitudes in the range 
19.5-21.5, and that are strongly clustered towards the luminous core of NGC~3603. 
However, if these are located at the distance of the cluster, they are far too luminous to be normal BDs.
We argue that it is unlikely that these objects are either artifacts of our dataset, normal field BDs/M-type giants 
or extra-galactic contaminants and, therefore, might represent a new class of stars 
having the effective temperatures of BDs but with luminosities of more massive stars. 
We explore the interesting scenario in which these objects would be normal 
stars that have recently tidally ingested a Hot Jupiter, the remnants of which are providing a short-lived
extended photosphere to the central star. In this case, we would expect them to show the signature of fast rotation. 

\end{abstract}


\section{Introduction \label{intro}}

Understanding the origin of brown dwarfs (BDs) is an important component of 
the theory of star formation and evolution that still remains somewhat mysterious \citep{Whi07}. 
Moreover, recent observations \citep{Gou09} suggest that old substellar field objects 
may be more common than previously assumed, opening again the debate on whether 
BDs are ubiquitous and occur in sufficient numbers to account for 
an appreciable amount of dark matter in the Galaxy \citep[e.g.,][]{Tin99}.  

Canonical approaches to these research fields include surveys for BDs in the field, studies of the sub-stellar initial 
mass function (IMF) in young clusters and associations under different star forming conditions, 
measurements of the disk frequency in the very low-mass regime and investigations of the mass accretion 
process near and below the substellar boundary \citep[see, e.g.,][]{Jay03}. 
However, these overall approaches ultimately rely on the spectral classification of low-mass stars and BDs and, hence, 
on optical/near-infrared (IR) spectroscopy \citep[see, e.g.,][]{McL03,Geb02}, which has proved 
impossible for very faint distant BDs with current telescopes. 
Thus, the studies conducted so far are limited to galactic clusters and star forming regions within a few kpc from the Sun, 
while the substellar population of extreme environments, such as the outer/center of the Galaxy or nearby galaxies, are basically unexplored. 
Such environments are particularly important for low-mass star/BD formation studies because they present properties 
(stellar density, total cluster mass, number of ionizing OB stars, metallicity, etc.) very different from nearby stellar populations 
and allow us to probe/discard the numerous scenarios proposed for BD formation \citep[e.g.,][]{Joe05}.

To overcome these limitations, new BD classification methods have been devised on the basis of near-IR broad-band \citep[see, e.g.,][and references therein]{Leg10,Sch10} 
and narrow-band  photometry that targets unique molecular features of L and T dwarfs, in particular H$_2$O and CH$_4$ bands \citep[see, e.g.,][]{Gor03,Mai04}.
These techniques are equivalent to extremely low resolution spectroscopy and can be confidently 
applied for statistical purposes, e.g. to detect and classify \emph{bona fide} 
very low-mass stars and BDs in large imaging surveys. However, such samples 
are inevitably affected by spectral contamination, since ground-based measurements of the depth of water bands
are complicated by the presence of variable water bands in the Earth's atmosphere itself. 
Thus, these kinds of studies are more efficiently conducted from space and, indeed, HST/NICMOS photometry has been 
already used to measure the strength of water bands for BD spectral classification purposes \citep{Naj00,And06}. 
However, NICMOS's field of view and sensitivity are still limited \citep{Tho98}. 
The new Wide Field Camera 3 (WFC3) on board the HST is very sensitive at both optical and IR wavelengths 
and brings unique capabilities to star formation studies. Specifically, the WFC3 is equipped with a set of four near-IR narrow-band filters 
spanning the J and H bands and covering H$_2$O molecular features useful for BD 
characterization, overcoming the contamination by atmospheric water bands affecting ground-based measurements 
and exceeding the sensitivity and field coverage of NICMOS.

In this paper we present WFC3/IR observations, including narrow-band imaging in the H$_2$O bands, of the young massive cluster in the core of NGC~3603. 
NGC~3603 ($RA_{\rm J2000} \approx 11^{\rm h} 15^{\rm m} 7^{\rm s}$, $Dec_{J2000}  \approx -61\degr 15\arcmin 30\arcsec $) 
is a galactic giant HII region at a distance of 6-7~kpc \citep[][and references therein]{Mel08}.  
The young \citep[1-20~Myrs,][]{Bec10} compact stellar cluster (HD97950 or NGC~3603YC) lies at the core of this region 
and has long been the center of attention for the relatively numerous population of massive stars (3 WNL, 6 O3-type stars and numerous late O-type stars). 
Together, these regions have a bolometric luminosity of 100 times that of the Orion cluster and 0.1 times that of NGC~2070 in the 30~Doradus 
complex in the Large Magellanic Cloud and a total mass in excess of 10$^4$~M$_{\odot}$ \citep[see][and references therein]{Har08}. 
These overall characteristics make NGC 3603 the closest small-scale resolved prototype of the starburst clusters commonly seen in active galaxies and 
the first natural place to probe the substellar population of an environment very different from those observed so far. 
The pioneering observations presented in this paper represent the ``first look'' at a star-burst cluster 
in the H$_2$O bands and the first attempt to ``see'' into its BD population. 

This paper is organized as follows: Sect.~\ref{data} describes the WFC3 observations in NGC~3603 and the data reduction procedure. 
In Sect.~\ref{tool} we develop a method to identify and classify BDs on the basis of WFC3/IR colors in the H$_2$O bands, 
discuss its advantages and limitations in comparison with predictions from synthetic spectra of stars and BDs and 
use the observations in NGC~3603 to quantify the effect of uncertainties in interstellar extinction.
In Sect.~\ref{sel} and \ref{discuss} we apply our method to identify objects presenting H$_2$O absorption bands in the field of NGC~3603 
and discuss the surprising properties of the selected sample. The conclusions of this work are given in Sect.~\ref{conclu}.

\section{Observations and data reduction \label{data}}

NGC~3603 was observed between August and November 2009 with the WFC3 on board the refurbished HST 
in both the optical (UVIS) and infrared (IR) channels as part of the early release science (ERS) program. The analysis of the UVIS data-set has been 
presented in a previous paper dedicated to the characterization of the stellar population in NGC~3603 and 
to the study of its star formation history \citep{Bec10}. Here we focus on the WFC3/IR observations. 

\begin{figure*}
\includegraphics[angle=0,scale=0.7]{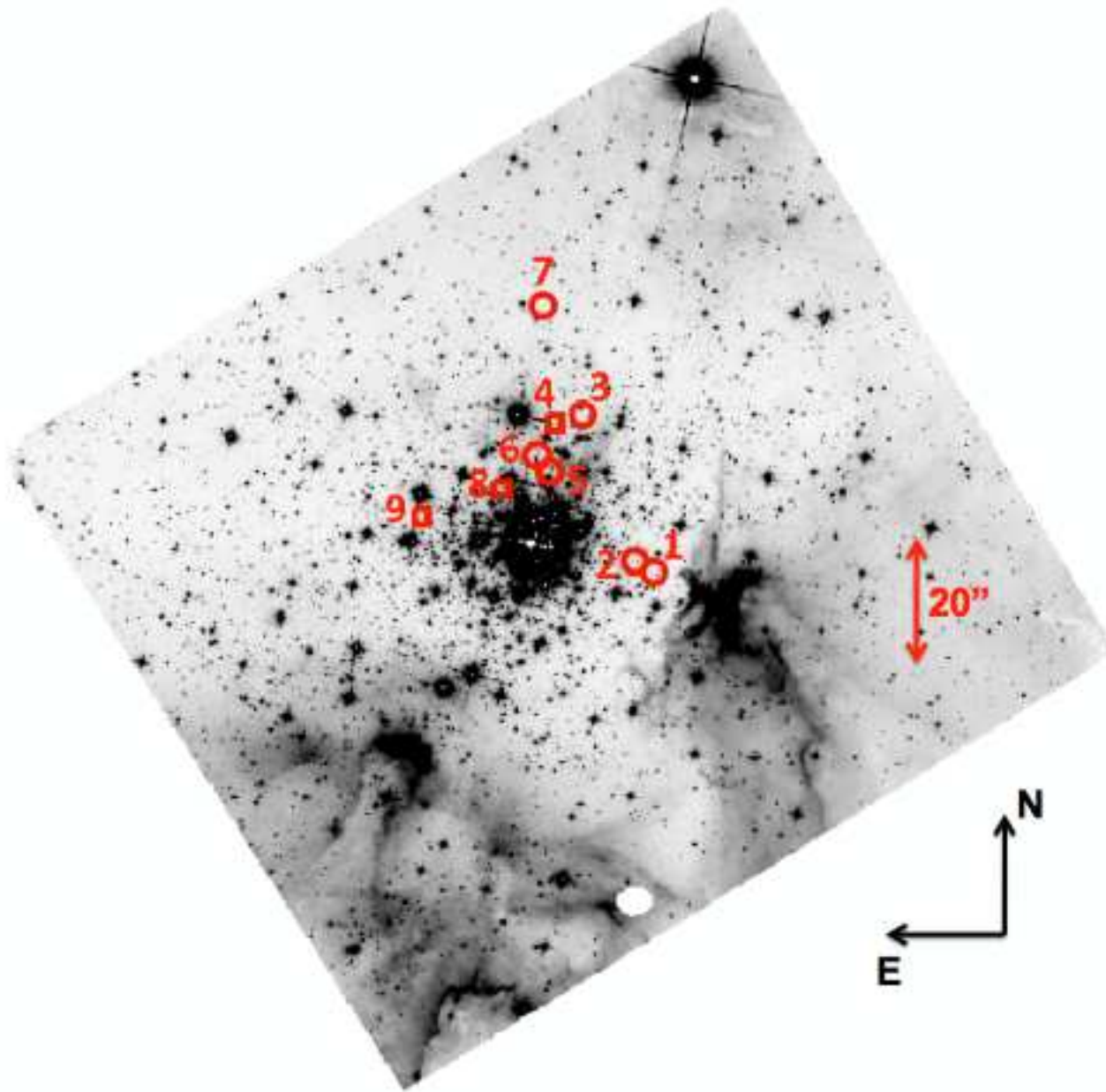}
\caption{WFC3 $2\arcmin \times 2\arcmin$ mosaic image of NGC~3603 in the F110W filter as obtained with the 
PyRAF/MULTIDRIZZLE package. The open symbols indicate the positions of the 9 objects showing water-vapor absorption 
discussed in Sect.~\ref{sel} and \ref{discuss} (squares indicate the 3 objects whose photometry might be contaminated by nearby bright/saturated stars). 
All of them are located within a radius of $\sim$30~arcsec from the cluster center (i.e. $\sim$1~pc at the cluster distance). 
Object ID numbers are as in Table~\ref{tab_BD}. The diffuse emission in the lower part of the image is due to scattered 
light from the reflection nebula surrounding the cluster  and to the nebular Paschen-$\beta$ emission line (1.2818~$\mu$m). \label{3603_im}}
\end{figure*}

The WFC3 IR detector is a single 1024$\times$1024 HgCdTe CCD 
with a total field of view of 123$\arcsec \times$136$\arcsec$ at a plate scale of 0$\farcs$13 per pixel. 
The observations in NGC~3603 were performed through the F110W (J-band) and F160W (H-band) broad-band filters, the 
F127M,  F139M and F153M medium-band filters, and the F128N narrow-band filter (see characteristics in Table~\ref{tab_filters}). 
For each filter, three images with approximately the same exposure time were taken with a constant $\sim$3.5~pixel 
dithering in order to allow for the removal of cosmic rays, 
hot pixels and other detector blemishes. 
The WFC3/IR detector is affected by persistence effects, which according to the WFC3 handbook 
are below the read noise in less than 15 minutes \citep[see Sect.~7.9.4. by][]{Dre10}. 
In order to ensure that persistence spots do not compromise the photometry, especially in the center of the cluster 
where there is a high concentration of bright OB stars, our observations were performed so that 
the time interval between two consecutive visits is always longer than 1.5 day (see Table~\ref{obs}). 
All observations were performed so that the core of NGC~3603 is roughly located 
at the center of the camera's field of view (FoV). In Figure~\ref{3603_im} we show a mosaic of the images in the F110W filter 
as obtained with the PyRAF/MULTIDRIZZLE package. 

\begin{figure}
\includegraphics[angle=0,scale=.35]{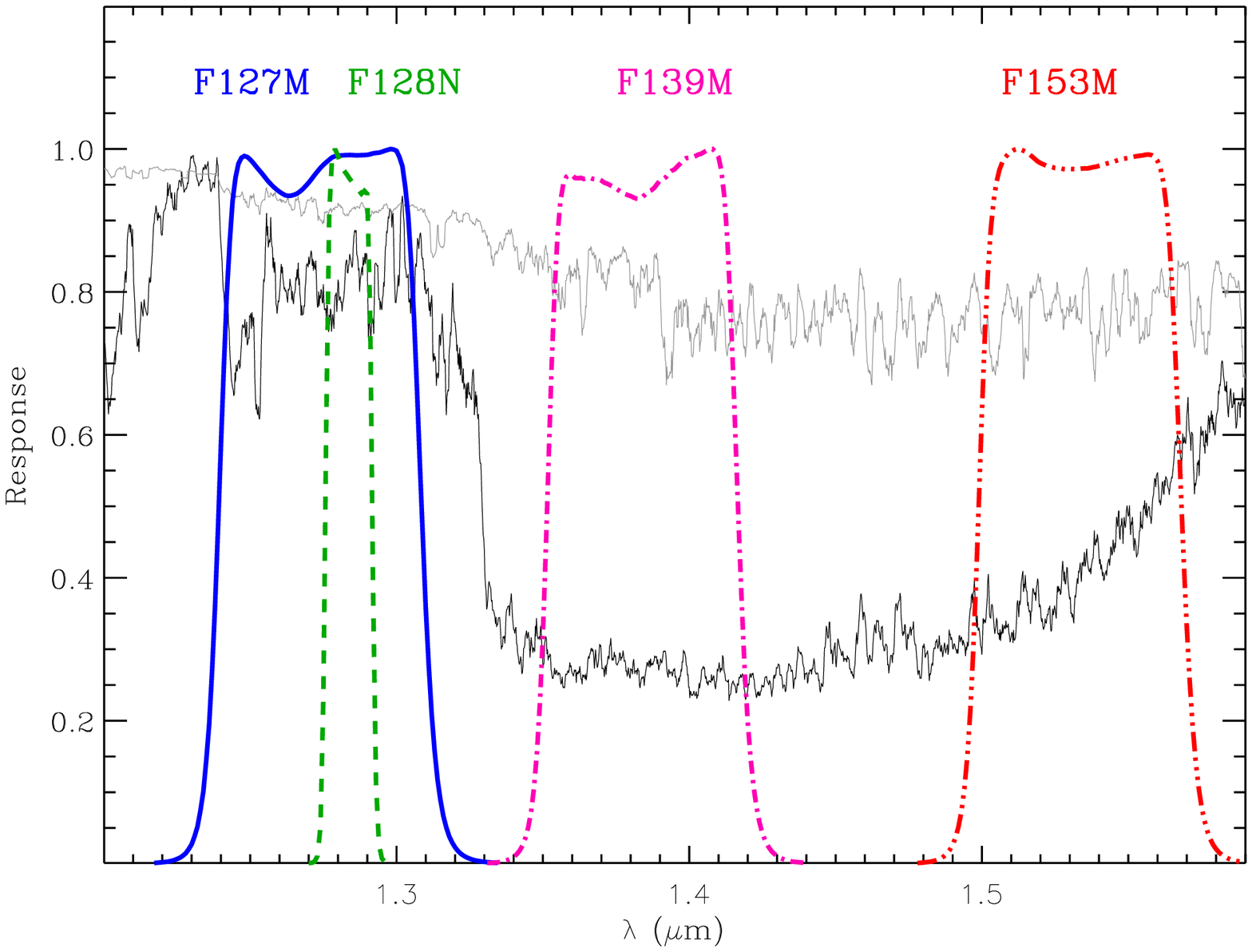}
\caption{The F127M, F128N, F139M and F153M WFC3/IR transmission bands. An example of a 
normalized AMES-Cond model BD spectrum \citep{Hau99} with 
$T_{\rm eff}=2000$~K and $\log g =4.0$ is over-plotted (black line). 
The broad H$_2$O absorption band centered at $\sim$1.4$\mu$m is clearly visible. 
For comparison, we plot the AMES-Cond model stellar spectrum \citep{Hau99} with 
$T_{\rm eff}=3500$~K (grey line), which does not show the H$_2$O feature.  \label{H02_filters}}
\end{figure}

The photometric analysis of the entire data-set was performed on the flat-fielded (FLT) images multiplied by the pixel-area maps to get uniformity 
in the measured counts of an object across the field. 
About 150 isolated and well-exposed stars were selected in every image over the entire FoV in order to 
properly model the point spread function (PSF) with the DAOPHOTII/PSF routine \citep{Ste87}. 
We find that a first order polynomial gives a good approximation to the PSF variation across the FoV.  
A first list of stars was generated by searching for objects above the 3$\sigma$ detection limit in each individual image 
and a preliminary PSF fitting was performed using DAOPHOTII/ALLSTAR. We then used DAOMATCH and DAOMASTER 
to match all stars in each chip, regardless of the filter, in order to get an accurate coordinate transformation between the frames. 
A master star list was created using stars detected in the F110W image (the deepest of the IR data-set) with the requirement that 
a star had to be detected in at least two of the three images in this filter.
We used the sharpness (sh) and chi square (chi) parameters given by ALLSTAR to remove spurious detections. 
The final catalogue was then obtained by rejecting residual spurious sources ($\sim$30\%) through visual inspection of the drizzled images. 
The master list was then used as an input for ALLFRAME \citep{Ste94}, which simultaneously determines the brightness of the 
stars in all frames while enforcing one set of centroids and one transformation between all images. All the magnitude 
values for each star were normalized to a reference frame and averaged together, and the photometric 
error was derived as the standard deviation of the repeated measurements. 
The final catalogue contains photometry in the F110W, F160W, F127M, F128N, F139M and F153M filters for 9693 stars.

The photometric calibration to the standard VEGAMAG photometric system was performed following \citet{Kal09}. 
A sample of bright isolated stars was used to correct the instrumental 
magnitudes to a fixed aperture of 0$\farcs$4. The magnitudes were then transformed into the VEGAMAG system 
by adopting the synthetic zero points for the WFC3/IR bands given by \citet{Kal09} (see their Table~5).

The WFC3/IR channel is affected by geometric distortion and a correction is necessary in order to properly derive the absolute positions 
of individual stars in each catalogue.
We used the distortion coefficients derived by \citet{Koz09} to obtain relative star positions that are corrected for distortion. We then used the stars in common
between our IR and the 2MASS catalogues to derive an astrometric solution and obtain the absolute R.A. and Dec. positions of our stars. 
We find a systematic residual of $\sim$0.3$\arcsec$ with respect to the 2MASS coordinates.

In Table~\ref{obs} we summarize the exposure times, saturation limits and limiting magnitudes at the 3$\sigma$ level in each filter. 
At the distance (6-7~kpc) and age (1-20~Myr) of NGC~3603 and in the absence of interstellar extinction, 
these limiting magnitudes correspond to cluster members with masses down to 0.15$-$0.2~M$_{\odot}$, according to the pre-main 
sequence (PMS) isochrones by \citet{Bar98} specifically calculated for the HST/WFC3 filters\footnote{Available at http://perso.ens-lyon.fr/isabelle.baraffe/}. 
Considering the typical extinction of A$_V$=5.5~mag toward the center of NGC~3603 (Sect.~\ref{ccd}), the mass limit would be $\sim$0.4~M$_\odot$. 
The large uncertainty on the distance to NGC~3603 \citep[$\sim$1~kpc or more;][]{Mel08} results in an uncertainty on our photometric limits of about 0.5~mag, 
i.e. $\sim$0.05~M$_{\odot}$ for cluster members. 
This means that our observations probe the cluster population down to the substellar boundary in the ideal case of no extinction and should be too shallow to detect sub-stellar members. 
However, as we will see in Sect.~\ref{sel} and \ref{discuss}, we identified a sample of very cool objects presenting H$_2$O 
absorption bands that are likely members of NGC~3603. Our method to identify objects with H$_2$O in absorption is based on the use of color indices (Sect.~\ref{tool}) and, 
hence, is not affected by the uncertainty on the cluster distance.

\begin{table}
\begin{center}
\caption{Characteristics of the WFC3/IR filters used in this paper: pivot wavelength ($\lambda_{P}$), band-width ($\Delta\lambda$) and 
photometric zero-point for the calibration to the VEGAMAG standard system. \label{tab_filters}}
\begin{tabular}{cccc}
\tableline
\tableline
Filter & {$\lambda_{P}$\tablenotemark{a}}  & {$\Delta\lambda$\tablenotemark{a}} & {Zp\tablenotemark{b}}  \\
       &  (nm)                             &   (nm)          &                        \\
\tableline
F110W  & 1153.4  & 443.0   & 26.07  \\
F127M  & 1274.0  &  68.8   & 23.69  \\
F128N  & 1283.2  &  15.9   & 21.98  \\
F139M  & 1383.8  &  64.3   & 23.42  \\
F153M  & 1532.2  &  68.5   & 23.21  \\
F160W  & 1536.9  & 268.3   & 24.70 \\
\tableline
\end{tabular}
\tablenotetext{a}{As defined in Sect.~7.5 of \citet{Dre10}.}
\tablenotetext{b}{From \citet{Kal09}.}
\end{center}
\end{table}

\begin{table}
\begin{center}
\begin{scriptsize}
\caption{Start time, total exposure time, saturation limit and limiting magnitude 
at 3$\sigma$ level in each WFC3/IR filter for the observations in NGC~3603.  \label{obs}}
\begin{tabular}{cccccc}
\tableline
\tableline
Filter  & Dithering & Star time    & T$_{exp}$   & Mag Sat.  &  Mag 3$\sigma$ \\
        &           & (Julian day) &   (sec)     &  	     &		      \\
\tableline          
F110W & 1 & 55070.226 &   3$\times$200   &   12.0   &  20.5 \\
      & 2 & 55070.291 & 		 &	    &	    \\
      & 3 & 55070.357 & 		 &	    &	    \\
\hline
F127M & 1 &  55070.241 &   3$\times$800   &   10.0   &  20.0 \\
      & 2 &  55070.306 &		  &	     &       \\
      & 3 &  55070.372 &		  &	     &       \\
\hline
F128N & 1 &  55070.251 &   3$\times$400   &   9.0    &  19.5 \\
      & 2 &  55070.315 &                  &          &       \\
      & 3 &  55070.382 &                  &          &       \\
\hline
F139M & 1 & 55070.453  &   3$\times$800   &   10.0   &  20.0 \\
      & 2 & 55070.519  &                  &          &       \\
      & 3 & 55070.586  &                  &          &       \\
\hline
F153M & 1 & 55070.424  &   3$\times$800   &   9.5    &  19.5 \\
      & 2 & 55070.491  &                  &          &       \\
      & 3 & 55070.557  &                  &          &       \\
\hline
F160W & 1 & 55070.229 &   3$\times$200   &   11.5   &  19.5 \\
      & 2 & 55070.294 & 		 &	    &	    \\
      & 3 & 55070.360 & 		 &	    &	    \\
\tableline
\end{tabular}
\end{scriptsize}
\end{center}
\end{table}

\section{Using WFC3 narrow-band photometry to characterize brown dwarfs \label{tool}}

The WFC3 is equipped with a set of four near-IR medium and narrow-band filters centered at 
1.27$\mu$m (F127M), 1.28$\mu$m (F128N), 1.39$\mu$m (F139M), and 1.53$\mu$m (F153M). 
The F139M filter probes the H$_2$O absorption band centered at $\sim$1.4$\mu$m 
while the F153M filter measures the edge of this band (Figure~\ref{H02_filters}) and is also sensitive to methane absorption. 
Other molecular absorption lines (CN, HCN, metal hydrides and alkali metals) characterize the 
spectrum of very cool objects (T$_{eff} \lesssim$3000~K) between 1.3 and 1.5$\mu$m, however 
H$_2$O is by far the strongest molecular absorber in this wavelength range \citep{Kir99,Kir05}. 
Because this H$_2$O feature is unique to very cool objects \citep[see Figure~1 in][]{Mai04}, the F139M and F153M filters 
can be used to identify very low-mass stars and BDs and estimate their approximate effective temperature by 
measuring the strength of the water absorption bands, provided the stellar continuum 
in this wavelength region is known. 
This information is easily obtained by combining measurements in the other two filters. 
Indeed, F127M can be used as the primary continuum filter for it covers a wavelength range 
relatively featureless in late-type objects; the only contamination expected in this wavelength 
range comes from the Paschen-$\beta$ line (1.2818$\mu$m), 
which is expected to be in emission in young accreting objects \citep[see, e.g.,][]{Fol01} 
and is present in nebulous regions as well  \citep[see, e.g.,][]{Ken98}. 
The narrow F128N filter, centered at 1.28$\mu$m with a bandwidth of 15.9~nm, 
is a good indicator of the possible contamination due to the Paschen-$\beta$ emission line. 
Thus, the flux ratio between the F139M and F153W bands
with respect to the continuum, corrected for Paschen-$\beta$ emission, 
gives a good measure of the strength of the water absorption bands. 

Our images are produced in units of electrons/sec and, hence, the counts scale
with the bandwidth of the filter. Thus, to compute the flux density (Flux) in the continuum and the water line, we
divided the corresponding counts (C) by the effective bandwidths of the filters:

\begin{equation}
cont = \frac{C_{F127M}-C_{F128N}}{\Delta\lambda_{F127M}-\Delta\lambda_{F128N}}\\
\end{equation}

\begin{equation}
Flux_{F139M} = \frac{C_{F139M}}{\Delta\lambda_{F139M}}\\
\end{equation}

\begin{equation}
Flux_{F153M} = \frac{C_{F153M}}{\Delta\lambda_{F153M}}
\end{equation}

where $\Delta\lambda_{F127M}$, $\Delta\lambda_{F128N}$, 
$\Delta\lambda_{F139M}$ and $\Delta\lambda_{F153M}$ are the 
bandwidths of the four filters reported in Table~\ref{tab_filters}. 
Equations 1-3 implicitly assume that the adopted filters have a rectangle response curve (top-hat filters), 
which is realistic in the case of WFC3/IR filters (Figure~\ref{H02_filters}).

We then convert the flux density in magnitude (mag) and define two narrow-band indices, 
I(139)=$mag_{cont}-mag_{F139M}$ and I(153)=$mag_{cont}-mag_{F153M}$, 
which are a measure of the water absorption and can then be used to identify very late-type 
objects and estimate their effective temperature.

\begin{figure*}
\includegraphics[angle=0,scale=0.75]{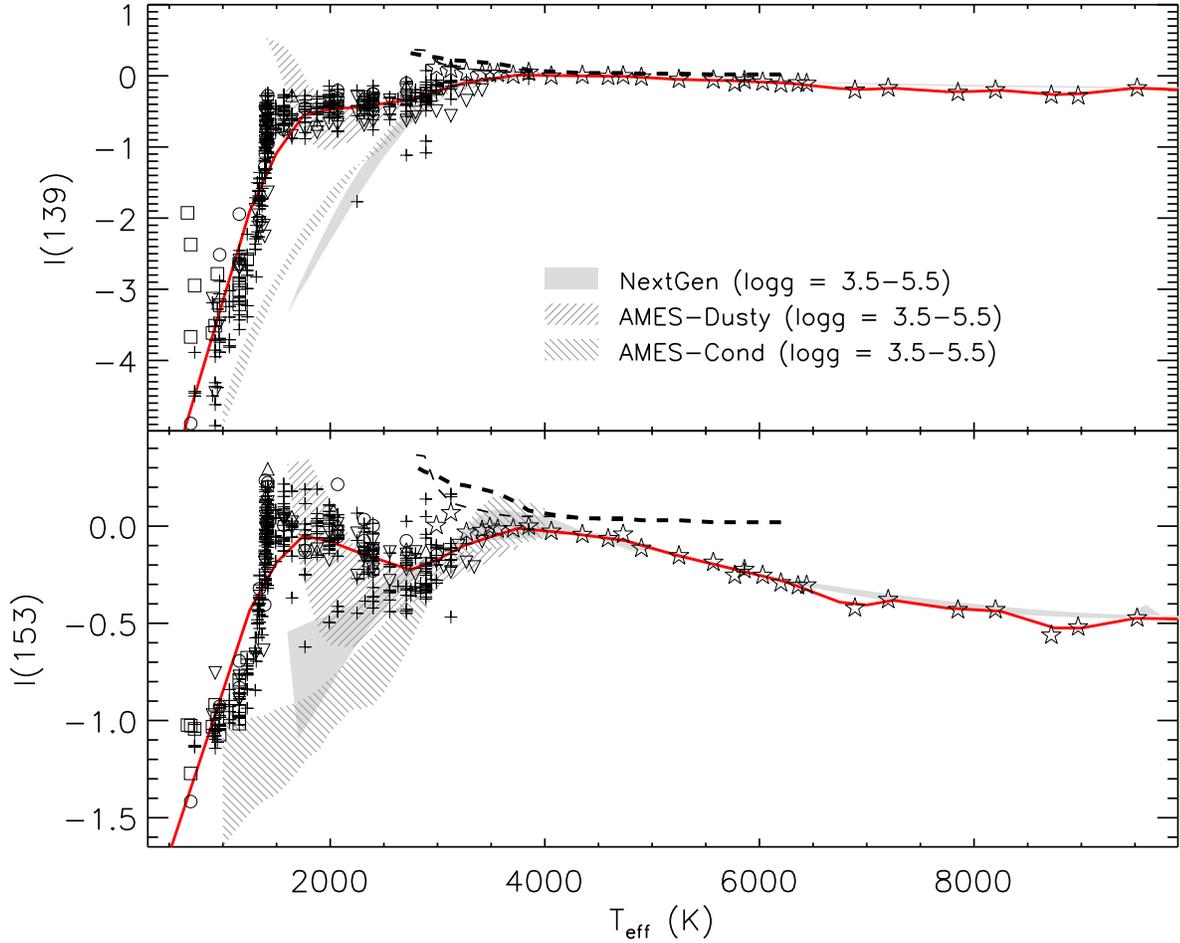}
\caption{I(139) and I(153) color indices as a function of the effective temperature for stars and BDs with 
actual spectra from the BPGS atlas (stars), the BDSS (circles), the SpeX libraries (plus symbols), \citet{Rei01} (triangles), 
\citet{Tes09} (upside down triangles) and B. Burningham (private communication, squares). 
The solid line is the as-measured calibration relation reported in Table~\ref{calib}. 
The grey-filled areas represent the relations derived from the model spectra, as indicated in the legend. 
The thin and thick dashed lines represent the calibration relation by \citet{Aar78} for 
giant and dwarf stars, respectively. \label{col_Teff}}
\end{figure*}

In order to verify this approach, we built an empirical relation between these color indices and the 
effective temperature by using the actual spectra of 552 stars and BDs available from the literature. 
These objects range in spectral type from O5 to T9 (T$_{eff}$=45000-600~K) 
and were drawn from the Bruzual-Persson-Gunn-Stryker (BPGS) atlas \citep{Str79,Gun83}, 
the Brown Dwarfs Spectroscopic Survey \citep[BDSS,][]{McL03}, the SpeX Prism Spectral Libraries (SpeX)\footnote{http://www.browndwarfs.org/spexprism/} 
and the spectroscopic databases for very low-mass stars and BDs 
by \citet{Rei01}, \citet{Tes09}, and B. Burningham (private communication). 
Using the WFC3/IR filter transmission curves and the STSDAS/Synphot synthetic photometry package in IRAF, 
we calculated the magnitudes of each object in each band by 
integrating the total flux in that band. Magnitudes are given in the VEGAMAG photometric system 
using the zero points for the WFC3/IR filters listed in Table~\ref{tab_filters}. 
Figure~\ref{col_Teff} shows the relation between the two water-band indices, derived from the 
observed spectra, and the effective temperatures (T$_{eff}$) of the 137 stars and BDs in our sample.
The figure shows that there is reasonable agreement between all of the sets of observations and 
the thick solid line represents the mean relation derived by averaging the measurements within steps of 250~K. 
Down to T$_{eff} \simeq$3500~K, both the I(139) and I(153) color indices have a mean value close to zero, 
indicating that no water is present in the spectrum. The onset of the water-vapor absorption feature 
is clearly seen below T$_{eff} \simeq$3500K, where both color indices become more and more negative as T$_{eff}$ decreases. 
This decrease is roughly monotonic in the range 3500$-$500K for I(139) and in the 
range 2500$-$500K for I(153) and, hence, these color indices can be used for BD spectral classification purposes.

We compared this empirical relation with expectations from model spectra of very-low mass stars and BDs. 
This can give further insight into the physical processes occurring in their atmosphere, 
which are still not clear \citep[see][and references therein]{Cha00}. 
We used the NextGen, AMES-Dusty and AMES-Cond model spectra 
which are, to our knowledge, the most complete grids of model atmospheres for late-type 
objects currently available. The NextGen models are dust-free model atmospheres \citep[see][]{Hau99}.  
The AMES-Dusty and AMES-Cond models \citep{All01} take into account the formation of condensed species, which 
depletes the gas phase of a number of molecular species and refractory elements, 
modifying significantly the atmospheric structure and, hence, the emergent spectrum.
While in the AMES-Dusty models the condensed species are included both in the equation of state
and in the opacity, taking into account dust scattering and absorption, 
in the AMES-Cond models the opacity of these condensates is ignored, in order 
to mimic a rapid gravitational settling of all grains below the photosphere. For further details about 
these grids of models we refer to \citet{Cha00}. 
The color indices of the model spectra were derived by folding them with the 
WFC3/IR filters, as was done with the actual spectra. The grey-filled areas in Figure~\ref{col_Teff} 
show the color vs. T$_{eff}$ dependency for the three sets of models. 
The comparison between the model predictions and the spectroscopic data shows that:

\begin{enumerate}

\item The relation between both our narrow-band color indices and T$_{eff}$ depends on stellar gravity, 
as indicated by the shaded areas in Figure~\ref{col_Teff}, which encompasses 
all the calibrations for models in the range $3.5\leq \log g \leq5.5$. This variation with $\log g$ is more 
significant for the I(153) color index, but in any case is well within the 
spread of the spectroscopic data;

\item the predictions by the three sets of models are in good agreement  with the actual spectra down to $\sim$3000K;

\item below $\sim$3000K the predictions of the three sets of models differ significantly from one another 
because the effect of dust grain formation in the atmospheres is treated differently, as explained above, and below 2000K 
they clearly disagree with the actual spectra. Data from the actual spectra show a linear decrease of both color indices with T$_{eff}$, 
more similar to the prediction by the AMES-Cond models. However, for both color indices 
this decrease is less steep than that predicted by the AMES-Cond models, mimicking an average 
behavior between the AMES-Cond and the AMES-Dusty models. In other words, 
the data support a scenario where both dust gravitational settling and scattering/absorption 
operate in the atmosphere of very cool objects and, hence, affect their spectrum.

\end{enumerate}

\citet{All01} suggest the use of the NextGen models for T$_{eff}>$2700K, the AMES-Dusty models for 1700K$<T_{eff}<$2700K, and 
the AMES-Cond models for T$_{eff}<$1400K. Between 1700 and 1400K, a new set of models (AMES-Settl) are being studied and 
are still in the experimental phase\footnote{See F. Allard web-page: http://perso.ens-lyon.fr/france.allard/}. 
Indeed, Figure~\ref{col_Teff} shows that this prescription roughly 
reproduces the trend obtained from the observed spectra. 
In particular, the AMES-Dusty models are roughly consistent with the I(139) and I(153) 
colors for 1700K$<T_{eff}<$2000K, while the AMES-Cond models underestimate both colors for T$_{eff}<$1400K 
for any stellar gravity in the range $3.5\leq \log g \leq5.5$. 
Thus, given the large uncertainties affecting model predictions in this temperature range, 
we conclude that the mean relation between the water-band indices and T$_{eff}$ derived from 
the observed spectra (solid line in Figure~\ref{col_Teff}) is more reliable for 
the identification and classification of BDs. 
We report this calibration relation in tabular form in Table~\ref{calib}; 
the uncertainties on the color indices are calculated as the dispersion of the data in each temperature bin.

We also show in Figure~\ref{col_Teff} the ground-based H$_2$O indices computed by \citet{Aar78} as a function 
of effective temperature for giant and dwarf stars. These indices were used for quite some time in constructing population
synthesis models for galaxies and clusters. It is then reassuring to see that our I(139) color index perfectly reproduces the 
trend of the \citet{Aar78} H$_2$O indices, while the I(153) only slightly underestimates them. 
Note that this agreement is not straightforward because the measurements were taken from different facilities 
(one ground-based, the other a satellite), different filters, different continuum estimates, etc.

\begin{table}
\begin{center}
\begin{scriptsize}
\caption{As-measured calibration relation between the water-band indices I(139) and I(153) and the effective temperature. \label{calib}}
\begin{tabular}{lccc}
\tableline
\tableline
Spec.$^\dag$ & T$_{eff}$  &  I(139) &  I(153) \\
Type         &     (K)    &                           &                           \\
\tableline
T9   &   500 &  -5.70$\pm$0.34  &  -1.68$\pm$0.05 \\
T8   &   750 &  -4.43$\pm$0.38  &  -1.27$\pm$0.01 \\
T6   &  1000 &  -3.17$\pm$0.41  &  -0.85$\pm$0.08 \\
T4.5 &  1250 &  -1.90$\pm$0.45  &  -0.43$\pm$0.15 \\
L6.5 &  1500 &  -1.08$\pm$0.33  &  -0.19$\pm$0.14 \\
L4.5 &  1750 &  -0.56$\pm$0.11  &  -0.05$\pm$0.08 \\
L2.5 &  2000 &  -0.49$\pm$0.09  &  -0.08$\pm$0.07 \\
L1.5 &  2250 &  -0.44$\pm$0.10  &  -0.13$\pm$0.07 \\
L0   &  2500 &  -0.39$\pm$0.10  &  -0.18$\pm$0.06 \\
M8   &  2750 &  -0.33$\pm$0.10  &  -0.23$\pm$0.06 \\
M6   &  3000 &  -0.21$\pm$0.09  &  -0.17$\pm$0.06 \\
M4   &  3250 &  -0.10$\pm$0.08  &  -0.10$\pm$0.06 \\
M2.5 &  3500 &  -0.04$\pm$0.05  &  -0.05$\pm$0.03 \\
M0.5 &  3750 &   0.01$\pm$0.02  &  -0.01$\pm$0.01 \\
K7.5 &  4000 &   0.01$\pm$0.01  &  -0.02$\pm$0.01 \\
\tableline
\end{tabular}
\end{scriptsize}
\end{center}
$^\dag$ Spectral types are inferred using the temperature scale by \citet{Vrb04} for L and T dwarfs, \citet{Luh03} for M dwarfs and \citet{Ken95} for K dwarfs.
\end{table}

\begin{figure*}
\includegraphics[angle=0,scale=0.75]{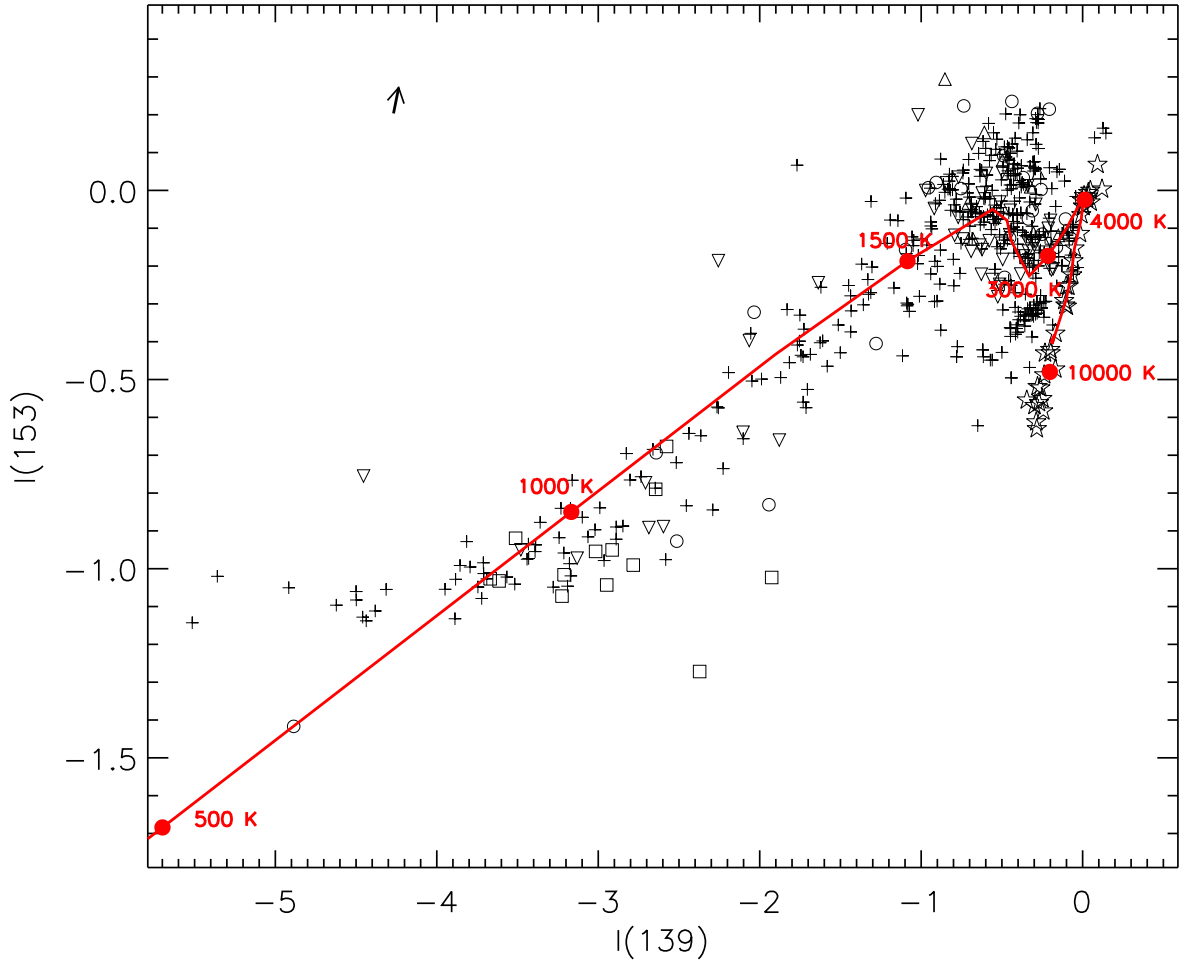}
\caption{I(139) vs. I(153) color-color diagram for stars and BDs with actual spectra (symbols are as in Figure~\ref{col_Teff}). 
The solid line is the as-measured calibration relation between the two color indexes reported in Table~\ref{calib}.
The arrow represents the A$_V$ =1 reddening vector.  \label{col_col}}
\end{figure*}

Figure~\ref{col_col} shows the I(139) vs. I(153) color-color plot 
derived from our calibration (thick solid line). This diagram shows that, when no information on the effective temperature of an object is available, 
WFC3 photometry in the four near-IR bands can be used to give an estimate of its T$_{eff}$ value. 
In particular, objects close to or below the sub-stellar limit (T$_{eff} \approx$3000K) 
are expected to be located in a region of this diagram well separated from that of more massive stars.  
We also used our BD reference sample to estimate the accuracy on the color-derived T$_{eff}$. 
We performed a linear fit of the spectroscopic T$_{eff}$ of the BDs in our reference sample as a function of the T$_{eff}$ predicted by the color$-$color relation. 
The computed RMS indicates that the two sets of measurements agree within $\sim$350~K. 
Thus, the accuracy of our method is sufficient for BD identification purposes, 
even though it provides only a rough estimate of their effective temperature. Two caveats must be taken into account:
i) our T$_{eff}$ estimate is based on the assumption that uncertainties on magnitude are not higher than 10\% in each of the four filters; 
ii) there has been a controversy in recent years over the spectroscopic effective temperatures scale for L and T dwarfs in that differences of the order of 250~K 
exist among different calibrations \citep[see, i.e.,][]{Bas00,Sch01,Sch02}.

Since our calibration relation is based on the use of near-IR filters
that are very close to each other in central wavelength, it is only slightly affected 
by uncertainties on interstellar extinction (Figure~\ref{col_col}). This makes our method particularly suitable to identifying 
BD candidates in young clusters, where extinction is normally highly variable. This issue is further discussed in Sect.~\ref{ccd}.

\subsection{Using the J-band as the primary continuum filter}

In this section we discuss the adoption of the WFC3/IR J-like bands as the primary continuum indicators 
when computing the I(139) and I(153) color indices.

As explained in Sect.~\ref{tool}, the use of the F128N filter allows us to correct the continuum emission for the possible contamination 
due to the Paschen-$\beta$ line, which is expected to be in emission in young accreting objects \citep[see, e.g.,][]{Fol01}. 
However, since F128N is a narrow-band filter ($\Delta\lambda =$13.54~nm), long exposure times are required to reach a photometric depth suitable for 
BD detection in distant star forming regions. Using the WFC3/IR-imaging exposure time 
calculator\footnote{http://etc.stsci.edu/webetc/mainPages/wfc3IRImagingETC.jsp}, 
we estimate that observations through the narrow F128N filter typically require an exposure time five times longer 
than for the three medium-band filters (F127M, F139M and F153M) to reach the same photometric depth. 
Thus, observations through the F128N filter are very time consuming and dominate the required observing time.

For  those objects which are not expected to accrete (e.g. field BDs) and are not located in nebular regions, 
the J-band  filter can be used as a continuum filter. 
The WFC3/IR camera is equipped with two J-like filters: F110W and F125W. 
In Table~\ref{calib_Jband} we provide the calibration relations between the water-band indices I(139) and 
I(153) and the effective temperature when the continuum is estimated using these two J-like filters.  
Note that, because both F110W and F125W are wide-band filters, the continuum estimates based 
on the use of these filters are slightly more sensitive to uncertainties in interstellar reddening.

\begin{table*}
\begin{center}
\caption{As-measured calibration relation between the water-band indices I(139) and 
I(153) and the effective temperature when using the F110W and F125W filters as continuum estimators. \label{calib_Jband}}
\begin{tabular}{lc|cc|cc}
\tableline
\tableline
&            &  Continuum filter F110W &  & Continuum filter F125W  &  \\
Spec.$^\dag$ & T$_{eff}$  &  I(139) &  I(153)  & I(139) &  I(153) \\
Type         &     (K)    &                           &                          \\
\tableline
T9   &   500   &  -5.26$\pm$0.58 & -1.25$\pm$0.23 &  -5.23$\pm$0.57 & -1.22$\pm$0.24 \\ 
T8   &   750   &  -4.10$\pm$0.54 & -0.93$\pm$0.22 &  -4.08$\pm$0.53 & -0.92$\pm$0.22 \\ 
T6   &  1000   &  -2.94$\pm$0.51 & -0.62$\pm$0.21 &  -2.94$\pm$0.50 & -0.62$\pm$0.20 \\ 
T4.5 &  1250   &  -1.77$\pm$0.47 & -0.31$\pm$0.20 &  -1.79$\pm$0.46 & -0.32$\pm$0.19 \\ 
L6.5 &  1500   &  -1.01$\pm$0.33 & -0.11$\pm$0.16 &  -1.04$\pm$0.32 & -0.14$\pm$0.14 \\ 
L4.5 &  1750   &  -0.50$\pm$0.11 &  0.02$\pm$0.10 &  -0.53$\pm$0.10 & -0.02$\pm$0.08 \\ 
L2.5 &  2000   &  -0.39$\pm$0.11 & -0.01$\pm$0.11 &  -0.44$\pm$0.09 & -0.05$\pm$0.08 \\ 
L1.5 &  2250   &  -0.34$\pm$0.14 & -0.03$\pm$0.13 &  -0.41$\pm$0.11 & -0.10$\pm$0.08 \\ 
L0   &  2500   &  -0.32$\pm$0.13 & -0.11$\pm$0.10 &  -0.37$\pm$0.10 & -0.16$\pm$0.07 \\ 
M8   &  2750   &  -0.30$\pm$0.11 & -0.20$\pm$0.08 &  -0.32$\pm$0.10 & -0.22$\pm$0.06 \\ 
M6   &  3000   &  -0.21$\pm$0.09 & -0.17$\pm$0.07 &  -0.22$\pm$0.09 & -0.18$\pm$0.06 \\ 
M4   &  3250   &  -0.11$\pm$0.08 & -0.11$\pm$0.07 &  -0.12$\pm$0.07 & -0.12$\pm$0.06 \\ 
M2.5 &  3500   &  -0.09$\pm$0.05 & -0.10$\pm$0.05 &  -0.09$\pm$0.04 & -0.10$\pm$0.04 \\ 
M0.5 &  3750   &  -0.07$\pm$0.03 & -0.10$\pm$0.03 &  -0.06$\pm$0.01 & -0.08$\pm$0.02 \\ 
K7.5 &  4000   &  -0.14$\pm$0.02 & -0.17$\pm$0.02 &  -0.09$\pm$0.01 & -0.11$\pm$0.01 \\ 
\tableline
\end{tabular}
\end{center}
$^\dag$ Spectral types are inferred using the temperature scale by \citet{Vrb04} 
for L and T dwarfs, \citet{Luh03} for M dwarfs and \citet{Ken95} for K dwarfs. 
\end{table*}

\subsection{Quantifying the effect of interstellar extinction \label{ccd}}

In this section we use the observations in NGC~3603 to quantify the effect of uncertainties 
in interstellar extinction on our BD classification method.

\begin{figure*}
\includegraphics[angle=0,scale=0.75]{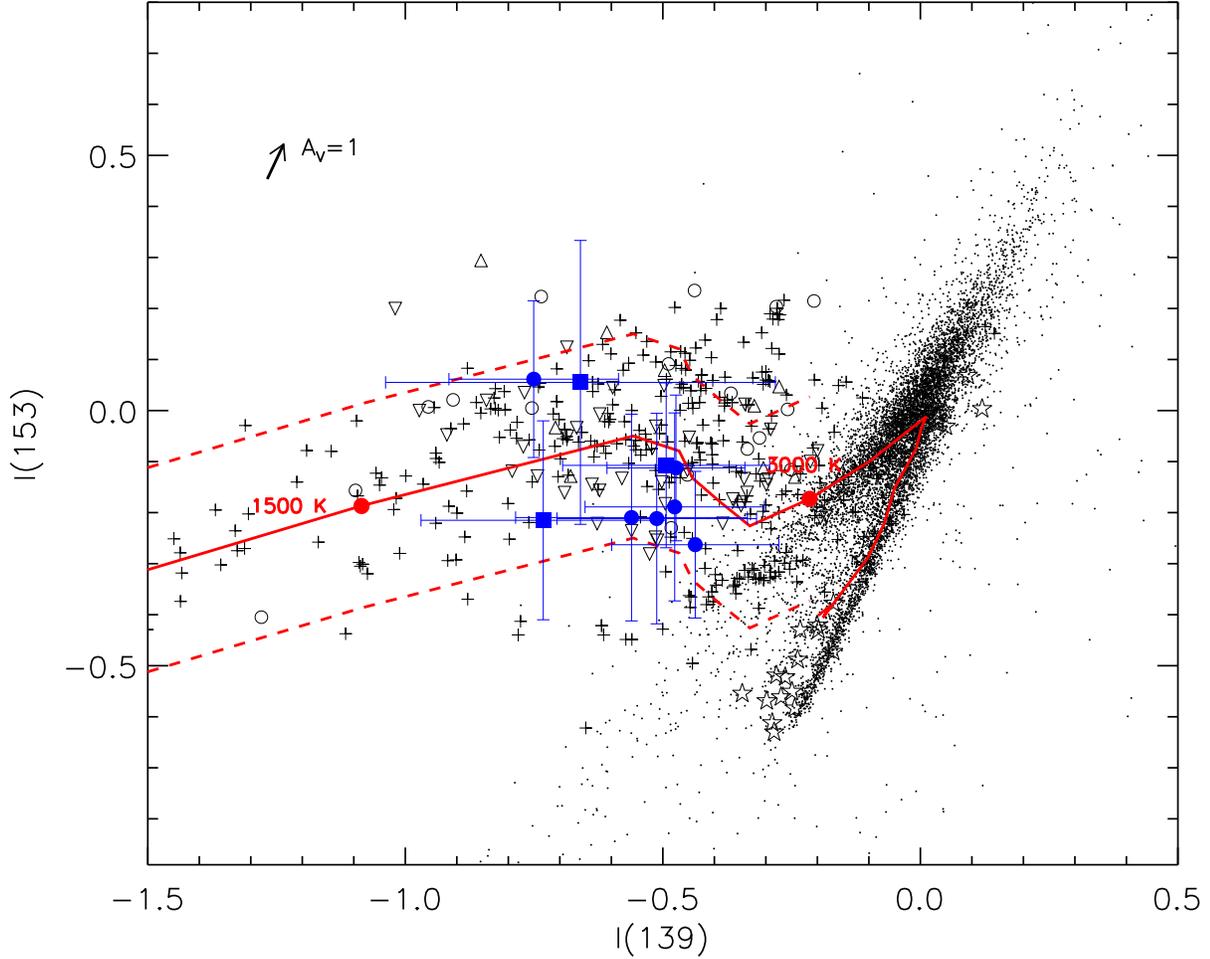}
\caption{Dereddened I(139) vs. I(153) color-color diagram for point-like sources in the field of NCG~3603 (small dots). 
The solid line is the calibration relation derived by us using stars and BDs with actual spectra; these spectroscopic data are represented as open symbols 
(as in Figure~\ref{col_Teff}) and the dashed lines represent their 1$\sigma$ dispersion in the substellar domain (T$_{eff} \lesssim$3000K). 
The larger filled symbols with error bars indicate the positions of the 9 objects showing water-vapor absorption 
(squares indicate the 3 objects whose photometry might be contaminated by nearby bright/saturated stars). 
The arrow represents the A$_V$=1 reddening vector.  \label{obs_cc}}
\end{figure*}

Figure~\ref{obs_cc} shows the dereddened I(139) vs. I(153) color-color 
diagram for point-like sources in the field of NCG~3603. 
In order to correct our magnitudes for interstellar extinction, we followed the study of differential reddening in NGC~3603 by \citet{Sun04}. 
These authors were able to map the variation of E(B-V) as a function of the distance from the cluster centre (see their Figure~5b) using multi-band HST
photometry of the bright massive stars (which are saturated in our images). They found that the value A$_V$=4.5 is representative of the very centre 
of the OB stars association, while they noticed an increase toward the external regions. 
Adopting the total-to-selective extinction ratio (R$_V$) value of 3.55, 
suggested by \citet{Sun04}, we estimate the value of A$_V$ =5.5 (i.e. A$_J \approx$1.5) 
to be representative of the mean visual extinction in the area sampled by our observations 
(from $\sim$10$\arcsec$  to $\sim$70$\arcsec$). We adopted this value of A$_V$ and the extinction law 
by \citet{Car89} to correct the magnitude of stars in our catalogue. 
As shown in Figure~\ref{obs_cc}, the dereddened photometry matches rather well our calibration relation, thus confirming the accuracy of our method. 

As mentioned in Sect.~\ref{tool}, since our calibration relation is based on the use of near-IR filters that are very close to each other in central wavelength, 
it is only slightly affected by uncertainties on interstellar reddening \citep[see also][]{Mai04}. 
The arrow in Figure~\ref{obs_cc} represents the reddening vector for A$_V$=1, 
which is the typical uncertainty in NGC~3603. This uncertainty in interstellar extinction translates the data points by 0.03~mag in  
I(139)  and 0.07~mag in I(153), i.e. within the photometric uncertainties.

Note also that, for large values of A$_V$,  R$_V$ depends on the intrinsic color of the star because 
of flux weighting through the different filter response functions \citep[][and references therein]{Wil63}. 
However, the fact that we compute colors from filters very close to each other in central wavelength makes this dependency moot.  

\begin{figure*}
\includegraphics[angle=0,scale=0.75]{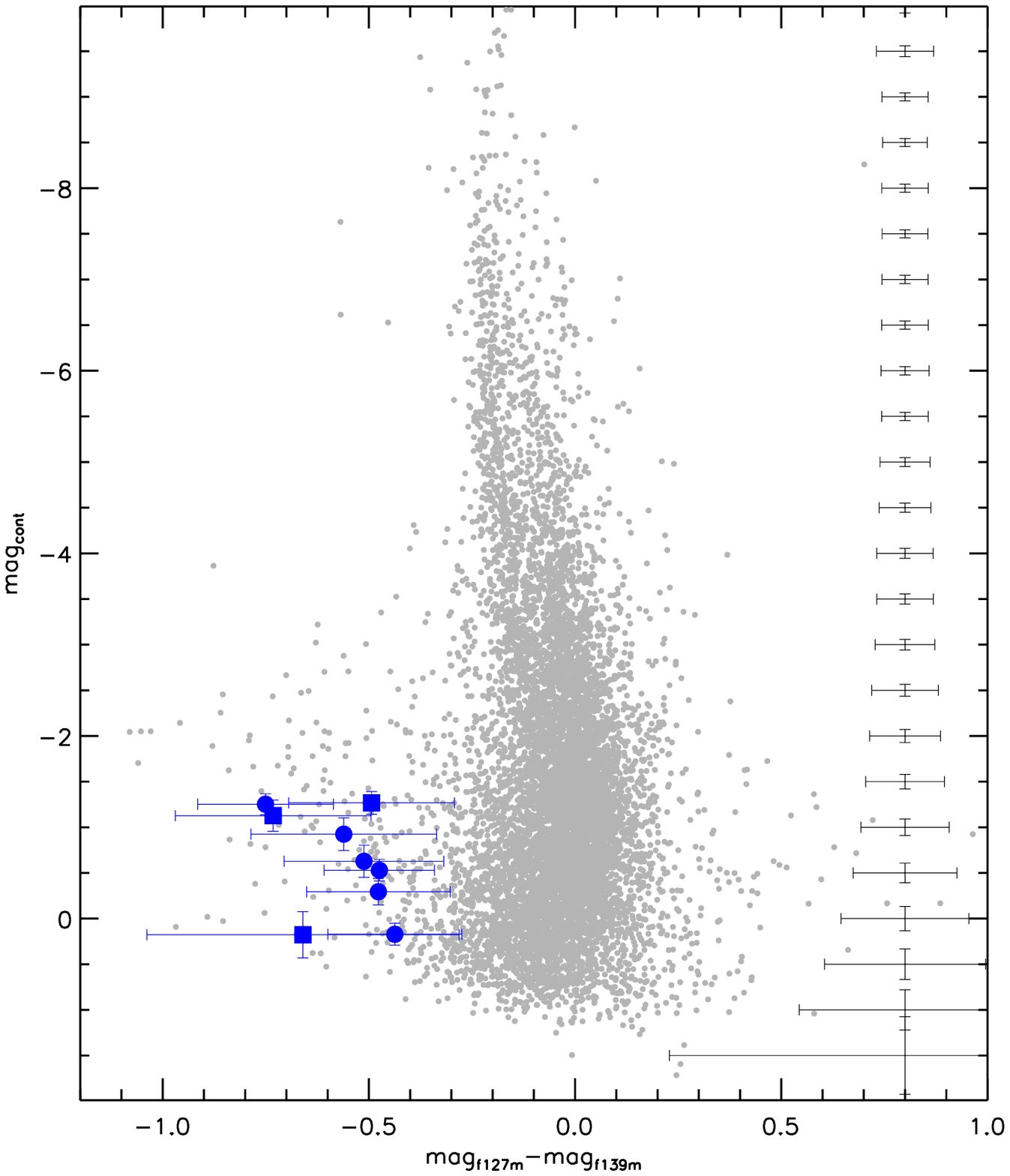}
\caption{Dereddened $mag_{cont}$ vs. $mag_{F127M}-mag_{F139M}$ color-magnitude diagram for point-like sources in the field of NCG~3603 (small dots). 
Average photometric uncertainties are reported on the right in steps of 0.5~mag. The bigger symbols with error bars indicate the positions of the 9 objects showing 
water-vapor absorption (squares indicate the 3 objects whose photometry might be contaminated by nearby bright/saturated stars).} 
\label{cmd_narrow}
\end{figure*}

\section{Selection of objects with water-vapor absorption in NGC~3603:  what are they?  \label{sel}}

As seen in Sect.~\ref{data}, our observations in NGC~3603 probe the cluster population down to the substellar boundary in the best case, i.e. in the absence of extinction, 
and hence we do not expect to detect any object presenting H$_2$O absorption bands belonging to this cluster. 
However, the dereddened I(139) vs. I(153) color-color 
diagram of the point-like sources in the field of NCG~3603 (Figure~\ref{obs_cc}) surprisingly 
shows a number of objects located in the region 
expected for BDs with significant absorption in the water bands (i.e. T$_{eff} \lesssim$3000K).

In order to systematically isolate this group of points, we first considered the calibration relation 
I(139) vs. I(153) in the substellar domain and calculated the dispersion ($\sigma_{spec}$) of the spectroscopic data used to derive this relation. 
We then selected all objects below T$_{eff} =$3000K having colors consistent with the 
calibration relation within 1$\sigma_{spec}$ and whose photometric uncertainties still place the object within 3$\sigma_{spec}$ of the mean calibration relation. 
These criteria take into account that, when measuring spectral line intensities, narrow-band photometry is necessarily less accurate than spectroscopy. 
We identified in this way 9 objects with absorption in the water bands, shown as filled circles/squares in Figure~\ref{obs_cc}. 
In Figure~\ref{cmd_narrow} we plot the position of these 9 objects on the dereddened $mag_{cont}$ vs. $mag_{F127M}-mag_{F139M}$ color-magnitude diagram. 
There are 9 objects which are singled out as interesting sources because they exhibit more negative  $mag_{F127M}-mag_{F139M}$ 
colors with respect to the remaining sources in the observed field, due to the 1.4$\mu$m  absorption band. 
Note that there are a number of objects with $mag_{F127M}-mag_{F139M}$ colors similar to these 9 objects which were not selected 
on the basis of Figure~\ref{obs_cc}, because they have larger photometric uncertainties on the I(139) and I(153) indices  and do not satisfy our selection criteria.

In Table~\ref{tab_BD} we list the observed WFC3/IR photometry for the 9 objects with water-vapor absorption 
and their derived effective temperatures, which lie in the range 1700-2200K. Their J-band observed magnitudes lie in the range 
19.5-21.5; assuming the distance modulus of NGC~3603 ($\sim$14~mag) and the average interstellar reddening in its direction (A$_J \approx$1.5), 
this range corresponds to J-band absolute magnitudes between 4.1 and 6.1 and to stellar luminosities in the range 0.6-4~L$_{\odot}$. 

\begin{table*}
\begin{center}
\begin{tiny}
\caption{Observed WFC3/IR photometry and estimated effective temperature for the 9 objects showing  water-vapor absorption.  \label{tab_BD}}
\begin{tabular}{ccccccccccc}
\tableline
\tableline
ID & RA~J2000)  & DEC~J2000  & m$_{F110W}$ & m$_{F127M}$ & m$_{F128N}$ & m$_{F139M}$ & m$_{F153M}$ & m$_{F160W}$ & T$_{eff}$ & Spectral$^\dag$ \\
    &   (hh:mm:ss)   &  (dd:mm:ss)  &                             &		         &			&			   &			       &			&  (K)          &   Type   \\
\tableline          
1           & 11:15:04.79  & -61:15:42.85  & 20.65$\pm$0.12 & 20.27$\pm$0.11  &  20.18$\pm$0.15  & 20.43$\pm$0.10  & 19.66$\pm$0.11  & 19.58$\pm$0.09  & 2100 & L2 \\  
2           & 11:15:04.93  & -61:15:42.75  & 20.43$\pm$0.13 & 19.98$\pm$0.14  &  20.08$\pm$0.17  & 20.13$\pm$0.08  & 19.35$\pm$0.11  & 19.23$\pm$0.09  & 2000 & L2.5 \\  
3           & 11:15:06.17  & -61:15:17.75  & 21.44$\pm$0.06 & 20.83$\pm$0.08  &  21.13$\pm$0.40  & 20.85$\pm$0.11  & 20.20$\pm$0.08  & 20.22$\pm$0.07  & 2200 & L1.5 \\ 
4$^\ddag$   & 11:15:06.82  & -61:15:19.26  & 19.73$\pm$0.23 & 19.48$\pm$0.14  &  19.55$\pm$0.14  & 19.85$\pm$0.16  & 18.85$\pm$0.09  & 18.67$\pm$0.12  & 1700 & L5 \\  
5           & 11:15:06.98  & -61:15:26.49  & 19.91$\pm$0.11 & 19.37$\pm$0.09  &  19.54$\pm$0.12  & 19.74$\pm$0.12  & 18.45$\pm$0.10  & 18.32$\pm$0.10  & 1700 & L5 \\   
6           & 11:15:07.00  & -61:15:25.98  & 20.46$\pm$0.15 & 19.66$\pm$0.14  &  19.64$\pm$0.17  & 19.88$\pm$0.14  & 19.05$\pm$0.09  & 18.79$\pm$0.09  & 1900 & L3.5 \\  
7           & 11:15:07.06  & -61:14:59.93  & 20.74$\pm$0.07 & 20.05$\pm$0.09  &  20.02$\pm$0.13  & 20.19$\pm$0.06  & 19.35$\pm$0.08  & 19.36$\pm$0.06  & 2100 & L2 \\  
8$^\ddag$   & 11:15:08.04  & -61:15:29.59  & 19.63$\pm$0.11 & 19.36$\pm$0.10  &  19.51$\pm$0.17  & 19.47$\pm$0.16  & 18.61$\pm$0.10  & 18.41$\pm$0.09  & 2100 & L2 \\  
9$^\ddag$   & 11:15:09.85  & -61:15:34.44  & 21.46$\pm$0.15 & 20.70$\pm$0.18  &  20.49$\pm$0.25  & 21.08$\pm$0.28  & 19.89$\pm$0.11  & 19.92$\pm$0.11  & 1700 & L5 \\  
\tableline
\end{tabular}
\end{tiny}
$^\dag$ Spectral types are inferred using the temperature scale by \citet{Vrb04}.\\
$^\ddag$ Possible photometric contamination from nearby bright/saturated stars.
\end{center}
\end{table*}

\begin{figure*}
\includegraphics[angle=0,scale=0.75]{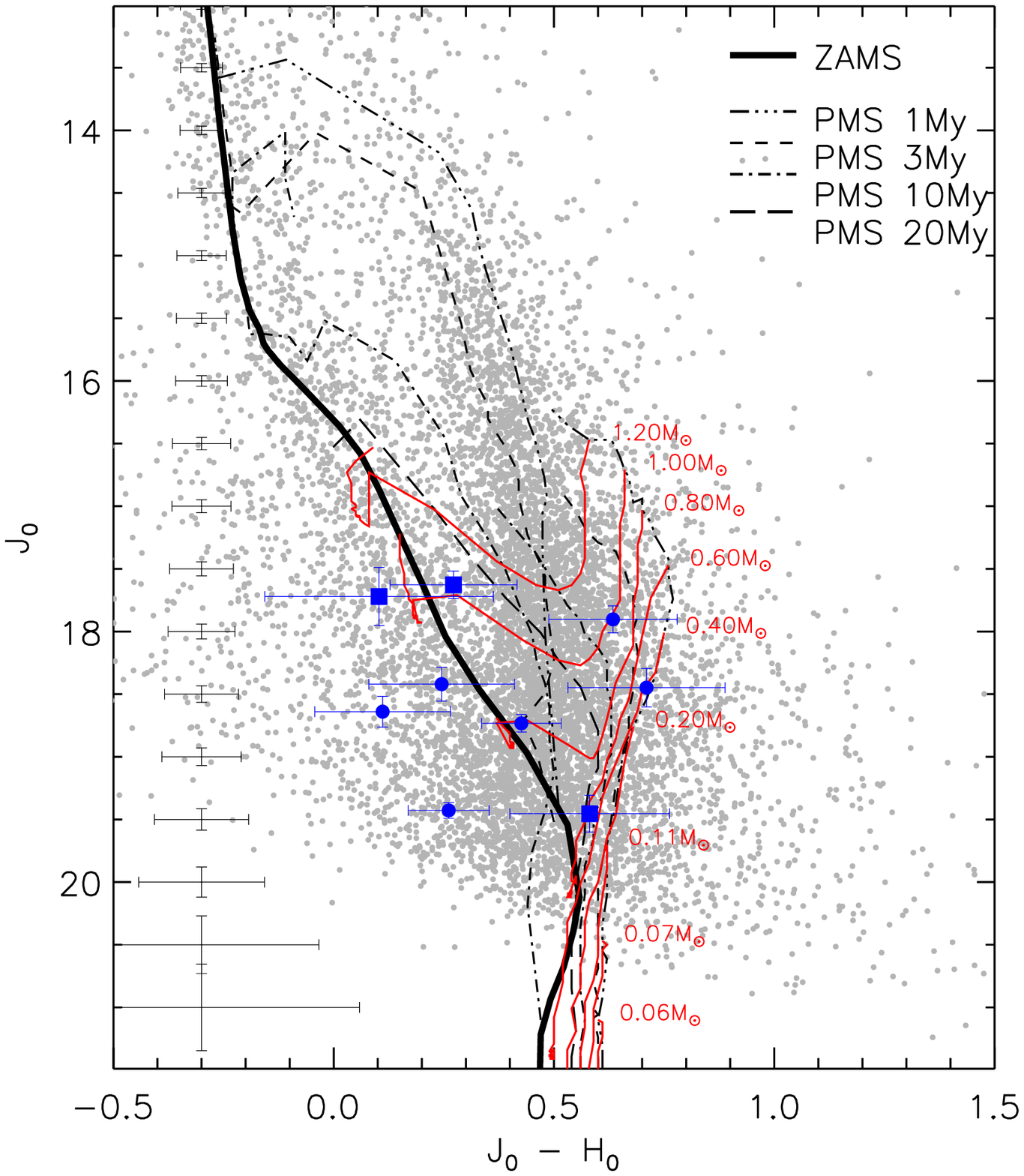}
\caption{Dereddened $J$ vs. $(J-H)$ color-magnitude diagram for point-like sources in the field of NCG~3603 (small dots). 
Average photometry uncertainties are reported on the left in steps of 0.5~mag.
The thicker solid line is the ZAMS of \citet{Mar08}, while the dashed lines are the PMS isochrones of \citet{Sie00} for masses 
M$>$1.4~M$_{\odot}$ and \citet{Bar98} for masses M$\leq$1.4~M$_{\odot}$ (see legend). The thiner solid lines are the evolutionary tracks by \citet{Bar98}. 
The bigger symbols with error bars indicate the positions of the 9 objects showing water-vapor absorption 
(squares indicate the 3 objects whose photometry might be contaminated by nearby bright/saturated stars). 
The masses implied for the objects showing water-vapor absorption bands are approximately in the range 0.2-1~M$_{\odot}$.}
\label{cmd}
\end{figure*}

Since the photometric sensitivity of our data does not allow us
to detect NGC~3603 members below 0.15-0.2~M$_{\odot}$ in the ideal case of no extinction, 
we now investigate what these objects are. 
Before exploring the various possibilities, we collect further hints on the nature of these objects making 
use of our observations in the broad-band F110W (J-band) and F160W (H-band) filters:

\begin{enumerate}

\item Figure~\ref{cmd} shows the dereddened $J$ vs. $(J-H)$ color-magnitude diagram (CMD) for the point-like objects detected in NGC~3603 
together with the zero age main sequence (ZAMS) of \citet{Mar08}, which roughly separates the locus expected for field main-sequence (MS) stars 
from the PMS star locus, and the PMS isochrones and evolutionary tracks by \citet{Bar98} and \citet{Sie00}, which take H$_2$O and other 
prominent near-IR absorption features into account\footnote{Note that isochrones have been corrected for a constant color term of $J-H$=0.18~mag 
to account for the difference between the standard $JH$ Johnson's filter and the WFC3 F110W and F160W filters.}. 
The position of the 9 objects with water absorption in this CMD indicates that their luminosity is comparable to that of PMS/MS stars 
with masses between $\sim 0.2-1~$M$_{\odot}$, i.e.  much higher than expected for PMS objects with a surface temperature in the range 1700-2200K. 
Moreover, only a few of them have $(J-H)$ colors typical of such cool PMS objects, while the majority appear to be bluer and, hence, older.
While the over-luminosity of these 9 objects has no straightforward explanation (see Sect.~\ref{discuss}), their  peculiar $(J-H)$ colors 
can be partly explained considering the star formation history of NGC~3603. 
Indeed, the analysis by \citet{Bec10} focused on stars with H$\alpha$ excess emission, a robust indicator of youth \citep[e.g.,][]{Whi03}, 
found that star formation in and around the cluster has been ongoing for at least 10-20 Myr; 2/3 of the 
young population have ages from 1 to 10 Myr with a median value of 3 Myr according to their position on the CMD, 
while 1/3 of them lie at or near the ZAMS, thus suggesting a considerably higher age, of order of 20-30 Myr. 
The 9 objects with water-vapor absorption appear to follow the same behavior, being spread between the PMS region of the CMD and the ZAMS. 
Although the statistic of this sample is hardly significant, we note that 
percentages are reversed with respect to the H$\alpha$ emitters population, i.e. $\sim$2/3 of the objects with absorption in the water bands 
lie near the ZAMS and only $\sim$1/3 (i.e. 2-3 objects) are located in the PMS region;

\item The spatial distribution of the 9 objects with water-vapor absorption appears 
to be strongly clustered towards the luminous core of NGC~3603 (see Figure~\ref{3603_im}). 
We attempt to use a Kolmogorov-Smirnov (KS) test to check whether the radial distribution 
of these 9 objects with respect to the cluster center is consistent with that of the cluster members 
\citep[see][for a detailed description of this test]{Bec10}. 
The small number statistic makes the results of this test not highly reliable; however, the general indication is that 
these 9 objects are more centrally concentrated than the field population, 
mimicking the radial distribution of the young stars belonging to the cluster. 
We also noticed that all 9 objects are preferentially located on the North side of NGC~3603 (Figure~\ref{3603_im}). 
The presence of pillars in the South region suggests a higher gas and dust density which could 
prevent us from detecting faint objects on this side of the cluster. 
A high resolution extinction map of NGC~3603 has been recently computed (X. Pang \& A. Pasquali, private communication) 
using WFC3 narrow-band imaging in the H$\alpha$ \citep{Bec10} and Pa$\beta$ lines (this work). 
The very preliminary result of this extinction study does not show a significant difference in reddening between the north and south regions of the cluster 
for the depth probed by the H$\alpha$ and Pa$\beta$ lines. However, the higher gas density in the South region can shield circumstellar disks from the radiation field of NGC~3603 more efficiently. 
This implies that circumstellar disks in the North and South region of the cluster might undergo different evolutionary paths.  
Indeed, as we well see in Sect.~\ref{discuss}, the observed properties of our 9 objects may be related to a specific disk dispersal mechanism, i.e. planet formation.
 
\end{enumerate}

Further clues as to the evolutionary phase of these 9 objects were 
obtained from optical broad-band and narrow-band H$\alpha$ photometry, 
which would allow us to put constraints on the mass accretion process typical of very young objects \citep[see, e.g.,][]{DeM10}. 
We cross-matched our WFC3/IR catalog with the optical WFC3/UVIS catalog of sources in NCG~3603 
by \citet{Bec10}. However, the 9 objects in exam are either not detected in the optical or do not have sufficient 
photometric accuracy to apply the method by \citet{DeM10}, which requires detection in the three $V$, $I$ and H$\alpha$ bands 
to determine the mass accretion rate of PMS objects.

We also searched for previous ground- and space-based observations of these 9 objects that could provide some hint on their variability. 

Two  ground-based imaging surveys of NCG~3603  are available to date: 
the $JHK$ survey conducted with  ISAAC@ESO-VLT by \citet{Bra99}  \citep[see also][]{And05}  and the deep, high angular resolution $JHKL'$ 
survey conducted with NAOS-CONICA@ESO-VLT by \citet{Har08}. 
The spatial resolution and sensitivity of the ISAAC data is 2-3 times poorer than for the WFC3 data. Thus, we could not retrieve accurate 
photometry of our 9 objects, which are located very close to the cluster center. 
However, all 9 objects are visible, though very faint, in $K$-band images. In the $J$ and $H$-band images, only 3 of them are detected with sufficient 
accuracy  to attempt photometry and the magnitudes we retrieved are consistent with the WFC3 ones within the photometric uncertainties. 
Only 3 of our 9 objects (namely ID~5, ID~6 and ID~8) are located in the  central 28$\arcsec \times $28$\arcsec$ 
area of NGC~3603 observed with NAOS-CONICA. Moreover, these observations have a detection limit of J$\approx$20 \citep[see Figure~3-4 in][]{Har08}, 
while our 9 objects have magnitudes very close to this limit or fainter. Thus, the ground-based observations available to date 
for NGC~3603 do not complement our WFC3-IR dataset.

NGC~3603 has been observed both with ACS (Advanced Camera for Surveys) and NICMOS (Near Infrared Camera and Multi-Object Spectrometer) 
on board of the HST using several narrow and wide-band filters spanning the 0.22-1.6$\mu$m wavelength range. 
These  observations were performed on 2005 December 29 and 1997 September 24, respectively. 
We retrieved the multi-drizzled images of this dataset from the HST Legacy Archive\footnote{http://hla.stsci.edu/}. 
Our 9 objects are not detected in the NICMOS F110W and F160W images because of their limited photometric 
depth \citep[see also][]{Mof04}, but are detected in the ACS images in the F435W and F850LP filters and 
their magnitudes are reported in Table~\ref{tab_BD_2}. 

Recently, \citet{Roc10} presented a proper motion study of the core of NGC~3603 based on HST/Wide Field Planetary Camera 2 (WFPC2) observations obtained 10 years apart, respectively, in 1997 and 2007. 
Only 3 of the 9 objects with water absorption fall in the 34$\arcsec \times$34$\arcsec$ area observed with the Planetary Camera (PC) and none of them is detected, most likely because their 
visual magnitude is fainter than the detection limit (m$_{555} \lesssim$25) of the \citet{Roc10} dataset.

\begin{table}
\begin{center}
\begin{scriptsize}
\caption{Observed ACS/Wide Field Camera photometry for the 9 objects showing water-vapor absorption. \label{tab_BD_2}}
\begin{tabular}{ccc}
\tableline
\tableline
ID &  m$_{F435W}$ & m$_{F850LP}$   \\
   &              &                \\
\tableline          
1   &          -       			& 23.603$\pm$0.547   \\  
2   &          -       			& 23.243$\pm$0.465   \\  
3   & 25.863$\pm$0.797 	& 23.625$\pm$0.555   \\ 
4   &          -      			 & 22.437$\pm$0.323   \\  
5   & 27.597$\pm$1.839 	& 21.685$\pm$0.229   \\   
6   & 26.659$\pm$1.171 	& 21.820$\pm$0.244   \\  
7   & 28.229$\pm$2.453 	& 23.426$\pm$0.505   \\  
8   & 25.532$\pm$0.728 	& 22.036$\pm$0.282   \\  
9   &          -       			& 23.772$\pm$0.601   \\  
\tableline
\end{tabular}
\end{scriptsize}
\end{center}
\end{table}

\section{Discussion \label{discuss}}

\begin{figure*}
\includegraphics[angle=0,scale=0.7]{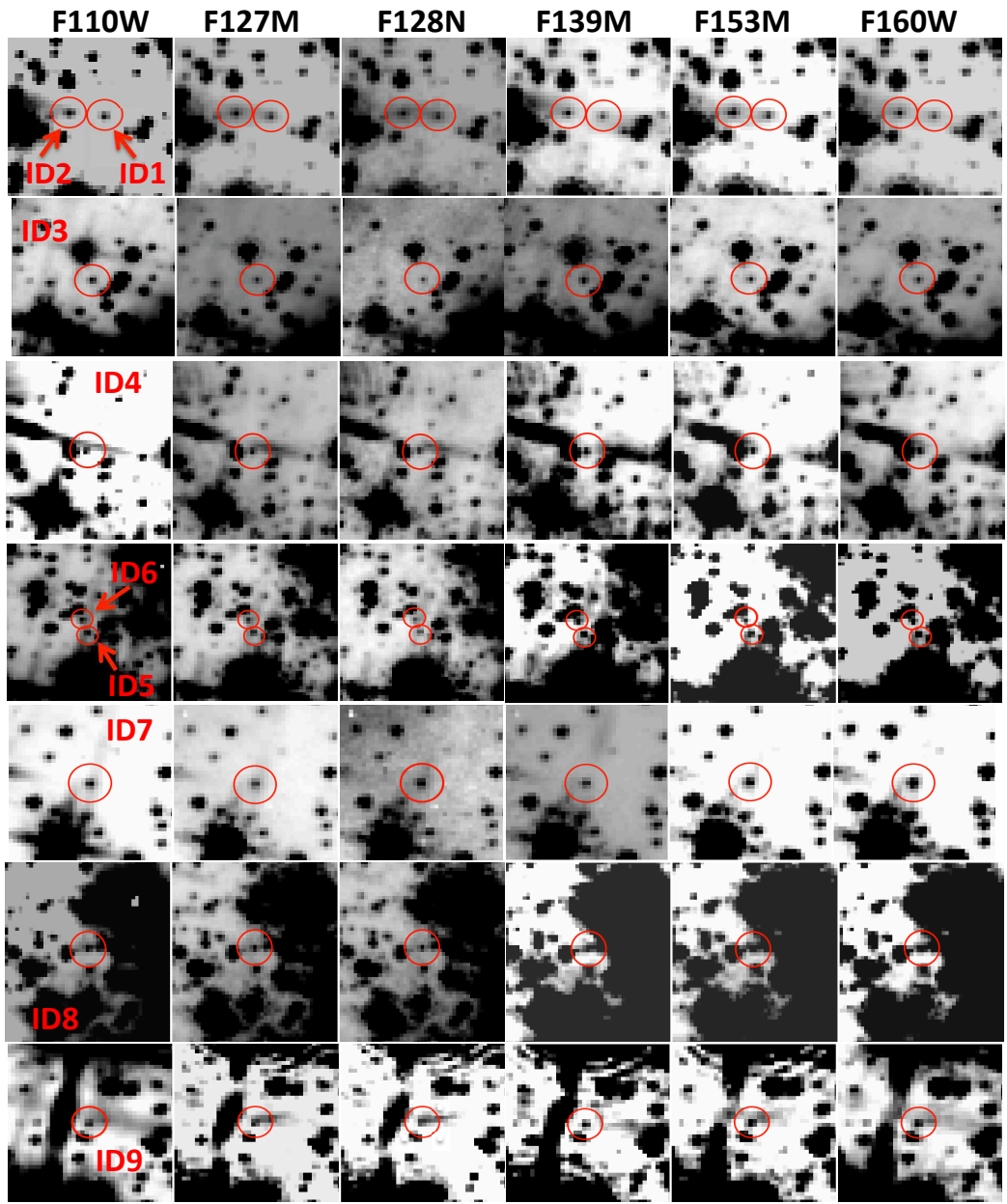}
\caption{Images of the 9 objects showing water-vapor absorption in each on the WFC3/IR filter used in this work, as indicated in the labels. 
Each snapshot covers an area of 5.2$\arcsec \times$5.2$\arcsec$; north is up and east to the left. \label{snapBD}}
\end{figure*}

Our exploratory narrow-band survey of NGC~3603 has led to the surprising identification of
9 objects considerably more luminous than BDs which nonetheless display
water-vapor absorption bands typical of BDs. Here we discuss possible explanations:

\begin{enumerate}

\item {\bf Contamination from field objects:} The most immediate explanation would be that the 9 objects 
presenting water-vapor absorption are field objects unrelated to NGC~3603. 
Under this hypothesis, these 9 objects could be either foreground field BDs or background early M-type super-giant stars, 
where the presence of H$_2$O absorption was confirmed by \citet{Tsu00}. 
Moreover, long period variables and late M-type supergiants in metal poor environments can have strong H$_2$O
absorption and, hence, bluer $J-H$ colors than expected for their T$_{eff}$ \citep[see, e.g.,][]{Eli85,Leb10}.  
All these hypotheses would explain why these objects appear over-luminous with respect to cluster BDs and why they are mainly located on the ZAMS 
in the CMD (Figure~\ref{cmd}), but on the other hand there are a number of objections
to it: Êa) there is no explanation for the concentration of the
contaminants around the center of NGC~3603 (Figure~\ref{3603_im}); 
b) metallicity studies in the direction of NGC~3603 confirm that the typical Galactic solar metallicity (z = 0.020) 
is appropriate for this region \citep{Sma02,Pei07,Leb08}; 
c) the \citet{Rob03} model of the Galactic stellar distribution does not predict any M-type super-giants in the 0.002~$deg^2$ area observed towards NGC~3603; 
d) using  the magnitude limits given in Table~\ref{obs}, the \citet{Cru07} luminosity function, and  
the distribution of L and T dwarfs in the Galaxy by \citet{Rya05}, we
find a predicted count in our field of 0.16 L-dwarfs for no extinction and 0.04 if the line-of-sight  
extinction is A$_V$=5.5~mag (Sect.~\ref{ccd}). However, Galactic counts models 
become more and more inaccurate towards the very low-mass regime and the predicted number depends on the adopted model. 
On the other hand, using other models \citep{DAn99,Rob03} would also result in no more than 1 or 2 LT dwarfs in this field. 
Thus, we do not expect to pick up any appreciable number 
of interloping field LT-dwarfs in the observed area towards NGC~3603. 
For all these reasons, Galactic field stars do not appear to be a major source of contamination 
of our sample of objects with water-vapor absorption. 

\item {\bf Extra-galactic contamination:} It is generally very hard to distinguish between 
an extended sources and background compact galaxies on the basis of broad-band colors alone \citep[see, e.g.,][]{Yan08}. 
Our sample of water-vapor absorbers might then be contaminated by extra-galactic objects. 
From the most recent J-band galaxy number counts \citep[see Table~1 and Figure~17 by][]{Cri09}, 
the expected number of background galaxies with 19.5$\lesssim J \lesssim$21.5, 
i.e. the magnitude range of the 9 objects under examination, in the observed $\sim$0.002~$deg^2$ area is $<$1. 
In addition, active galaxies usually show near-IR water lines in emission because of the presence 
of water vapor masers \citep[e.g.][]{Bra03}. Hence, they are not a likely source of contamination.

\item{\bf Artifacts:} Another hypothesis to check is whether there is some artifact in the data that has 
biased us towards finding cluster-centric sources. 
Our images are not significantly affected by instrumental effects, such as vignetting, optical ghosts or image
degradation moving away from the center, that could have caused bias in the selection. 
All 9 selected sources are well detected in our images, 
with a signal-to-noise ratio better than 5 in all 6 filters (Figure~\ref{snapBD}).
Moreover, all of them are visible, though very faint, in the ISAAC and ACS images (Sect.~\ref{sel}).
This demonstrates that they are not spurious detections. 
We have also checked whether the magnitudes of these 9 objects could be somehow affected by nearby bright/saturated stars. 
Also this possibility appears to be unlikely for various reasons: 
a) as explained in Sect.~\ref{data}, our observing strategy ensures that persistence spots 
due to saturated sources do not compromise the photometry; 
b) we excluded from our analysis the innermost 5\arcsec region of the cluster, where there is a high concentration of bright OB 
stars that are saturated; c) only 3 (see ID~4, ID~8 and ID~9 in Figure~\ref{snapBD}) out of the 9 sources are located 
within a few pixels of bright/saturated sources that might have contaminated their photometry, while all the others are not; 
these 3 objects are marked in Table~\ref{tab_BD} and in Figure~\ref{3603_im} and \ref{obs_cc}-\ref{cmd}.
We adopted PSF fitting magnitudes and local background estimates (see Sect.~\ref{data}), which should minimize 
photometric contamination even in those 3 cases. 
However, even in the most conservative case in which we exclude these 3 objects, we remain with 6 clustered objects showing water absorption, 
a number still hard to explain invoking field contamination, as seen above.

\item{\bf Binary systems:} The 9 objects could be BDs belonging to NGC~3603 which are detected because they are in binary systems and, hence, more luminous. 
Because the BD population of NGC~3603 is still not explored, we cannot predict the number of binary BDs in the cluster. 
However, they are to be expected, because the fraction of BDs comprised in binaries measured in the field is of the 
order of 15\% for a primary mass below 0.1~M$_{\odot}$ \citep{Lad06,Thi07,Bat09} and appears to be even higher 
in star forming regions and young clusters of age similar to NGC~3603  \citep[$\sim$20\%;][and references therein]{Kon07}. 
Moreover, \citet{Bur08} reported the detection of a field  BD, namely 2MASS~J09393548Ð2448279, presenting a highly inflated radius 
which can be reconciled with BD structure models of the same effective temperature (600-700~K) only assuming that this source is an unresolved, equal-mass binary.

A BD/BD or BD/planet pair would have a luminosity $\sim$0.75 mag brighter than a single BD at maximum, 
e.g. in the case of the two BDs of the pair having the same mass. 
According  to the isochrones by \citet{Bar98}, our 9 objects are about 2 mag brighter than what is expected for single BDs belonging to the cluster. 
This estimate is affected by our photometric uncertainties and also depends on two uncertain parameters: the distance to NGC~3603 and the interstellar extinction along the line of sight. 
The photometric uncertainties for our 9 objects are of the order of 0.10~mag  (see Table~\ref{tab_BD}). The cluster distance is known with 
an uncertainty of about 1~kpc \citep{Mel08}, i.e. $\sim$0.33~mag in distance modulus, while the uncertainty on visual extinction is $\Delta A_V \approx$1~mag (see Sect.~\ref{ccd}), 
i.e. $\Delta A_J \approx$0.28~mag. Altogether, these uncertainties might account for an uncertainty on absolute J magnitude of $\sim$0.7~mag. 
Adding this 0.7~mag to the $\sim$0.75~mag excess luminosity produced by a BD/BD pair, binary BDs would be at most 1.5 mag brighter than single BDs in the cluster. 
This difference is still  smaller than the 2~mag difference seen in the CMD (Figure~\ref{cmd}) between the position of our candidates and the expected position of BDs belonging to the cluster. 
Thus, the hypothesis of these objects being  BD/BD or BD/planet pairs appears unlikely.
To justify the observed luminosity, we should assume that they are star/BD pairs 
with the primary star having a mass between 0.3 and 0.6 M$_\odot$, so that the BD would be responsible for the observed water absorption 
while the primary star would be responsible for the observed luminosity. 
However, in this scenario, the luminosity of the primary star in the F139M and F153M filters
would suppress the BD contribution to our non-resolved photometry and, hence, we would not be able 
to detect the presence of water absorption bands.

\item {\bf A new class of objects (``Bloatars''):} The alternative hypothesis would be that these objects belong to 
a formerly unrecognized class of young stars with a surface temperature similar 
to those of BDs, but with a luminosity characteristic of much more massive stars ($\sim 0.2-1~$M$_{\odot}$). 
These stars appear to be ``bloated'', with radii far in excess of what standard PMS evolutionary theory would predict for objects of the same temperature, 
and so we will dub them ``Bloatars'' (bloated stars). 
All the hints we have collected on these 9 objects on the basis of our data support this hypothesis: 
their cool temperature, their higher luminosity with respect to field and/or cluster BDs, their clustered spatial distribution, 
their CMD similar to that of PMS stars in NCG~3603. 

\end{enumerate}

In the next sections we explore this possible and interesting scenario in more detail.

\subsection{How could ``Bloatars'' be formed? \label{timescale}} 

The data presented in this paper suggest that we have discovered a new class of PMS stars, designated as ``Bloatars'', 
with a surface temperature similar to those of BDs but with a luminosity characteristic of much more massive stars ($\sim$0.2-1~M$_{\odot}$). 
How could such stars be formed? Essentially, it would require a recent mass accretion event - preferably with added angular momentum. 
Two scenarios are possible. First, that the system was formed as a very close binary system, 
and that the secondary has overflowed its Roche Lobe. 
Second, the Bloatar has recently swallowed an orbiting hot Jupiter planet.
In both these scenarios, the merging is driven by tidal interaction, and the debris 
from the merger forms a common envelope which is optically-thick to the escape of 
photons from the primary star's photosphere, thus forming - for a brief period - a ``false photosphere'' 
with radius much larger, and temperature much lower, than the primary star.  
We will now investigate the likely parameters of such tidal mergers, 
and attempt to determine which scenario gives the most plausible results.

\subsubsection{Formation Rates and Timescales}

Planetary systems forming hot Jupiters are thought to form in two stages, an initial accretion period lasting a few Myr, 
with orbital migration driven by dynamical interaction with the protoplanetary disk, followed by a chaotic era 
lasting as long as 100~Myr in which interactions 
between the newly formed planets dominate \citep{Juric08}.

The age of the NGC~3603 cluster stars varies from zero up to more than 10~Myr. 
\citet{Bec10} show that the star formation rate has been increasing rapidly over the past few Myr, 
with the bulk of the stars forming over the past $\sim$2-3~Myr. Thus, the age of the cluster stars 
is comparable to the length of the first phase of evolution of the planetary system. Any planet which 
is going to merge with its star must have therefore already migrated to an orbit in which continued 
tidal interaction will cause it to inspiral in a timescale of less than about 1-3~Myr.

The inspiral time is determined by a dimensionless parameter $Q$ representing the ratio of the binding energy 
and the increase in binding energy over one orbital period. This is driven primarily by tidal dissipation 
in the satellite body, but may also be augmented by magnetic braking and spin-orbit alignment \citep{Barker09}. 
Typically, $10^5< Q < 10^6$ \citep{Peale99}. The inspiral timescale is given by \citep{Dobbs04,Levrard09}:

\begin{equation}
t_{\rm in} \sim {{2Q}\over{117n}} \left({M_*}\over{m_{\rm p}}\right) \left({a}\over{R_*} \right)^5
\end{equation}

where $n$ is the orbital angular frequency, the star has mass $M_*$ and radius $R_*$, and the planet 
has mass $m_{\rm p}$ and is orbiting with semi-major axis $a$. If we want the star to inspiral in $\sim 1$~Myr, 
then the hot Jupiter must initially be at a radius of order $0.02-0.03$~AU ($3-4\times10^{11}$cm) 
with an orbital period of the order of a day. Hot Jupiters with such short orbital timescales have 
indeed been observed, e.g. the eclipsing system WASP-18 \citep{Hellier09}. 
If the companion was not a planet, but a star in its own right, the infall 
timescale from this initial radius would be perhaps a factor of 100 times longer.

The number, $N_{B}$, of Bloatars we expect to see is $N_{B} \sim fN_{\rm C}{{t_{\rm B}}\over{t_{\rm C}}}$ 
where $N_{\rm C}$ is the number of stars in the cluster with mean age $t_{\rm C} \sim 1$~Myr, 
$f$ is the fraction of stars having hot Jupiters which can inspiral in the lifetime of the cluster, 
and $t_{\rm B}$ is the lifetime of the bloated phase, before the accreted envelope once again becomes 
optically thin to the escape of photons from the true stellar photosphere. 
According to the most recent estimates, though still uncertain, the value of $f$ is expected to be within the range 0.1-0.2 \citep[see, i.e.,][]{Cum08}, 
while $N_{\rm C}$ in NGC~3603 is of the order of $\sim 1000$ \citep{Bec10} and $N_{B}$ is $\sim 10$ according to this work. 
Thus, $t_{\rm B} \sim 5 \times 10^4$~yr. Bloatars must be short-lived, and therefore comparatively rare objects. 

\begin{table*}
\begin{center}
\caption{Expected effective temperature of the bloated photosphere (T$_{eff}$), according to Equation~\ref{Teff}, 
for different combinations of mass/temperature of the primary star (M$_\star$, T$_\star$) and mass/orbital radius 
(m$_p$, r$_p$) of the companion planet. \label{bloat_Teff}}
\begin{tabular}{cc|ccccc}
\tableline
    &   & m$_{p}$ (M$_\odot$)        & 0.001 &  0.01 & 0.001 & 0.0005  \\
    &   & r$_{p}$ (R$_\odot$)           & 0.14   &  0.14 & 0.25   & 0.14    \\
\tableline
M$_{\star}$ (M$_\odot$) & T$_\star$ (K)  & T$_{eff}$ (K) &  &  &  & \\
1.00  & 5920  &   & 3326  & 4755  & 2489  & 2977 \\
0.93  & 5610  &   & 3075  & 4393  & 2301  & 2752 \\
0.78  & 5150  &   & 2775  & 3958  & 2076  & 2485 \\
0.69  & 4640  &  & 2378  & 3388  & 1780  & 2130 \\
0.47  & 3920  &   & 1970  & 2794  & 1474  & 1765 \\
0.21 &  3120  &  & 1266  & 1775   & 948    & 1137 \\
\tableline
\end{tabular}
\end{center}
\end{table*}

\subsubsection{Orbital parameters \& Surface temperatures}

An inspiralling planet or companion star will start to be tidally disrupted when it comes to 
fill its Roche lobe. According to \citet{Paczynski71}, for mass ratios in the range 
$ 0 <  \left({m_{\rm p}}\over{M_*}\right) <0.8$ this will occur when the radius 
of the companion, $r_{\rm p}$, exceeds a critical radius, $R_{\rm cr}$, given by:

\begin{equation}
r_{\rm p} > R_{\rm cr} = 0.46224a \left({m_{\rm p}}\over{M_* + m_{\rm p}}\right)^{1/3}.  \label{rcrit}
\end{equation}

We assume that the tidally stripped debris from the companion form a fast-rotating common envelope 
with a radius $R_2 = (a+r_{\rm p})$. The stripped debris form a "false photosphere" to the star; if the tidal debris are optically thick, the effective radius 
of the parent star is $\sim R_2$ and therefore the effective temperature of the  "false photosphere" is given by:

\begin{equation}
T_{\rm eff} \sim T_*\left({R_*}\over{R_2}\right)^{1/2}  \label{Teff}
\end{equation}

Presumably, the effective temperature would be higher if viewed from the polar direction, because the common 
envelope would be ellipsoidal in shape. If the companion is a lower mass star, then nuclear burning is not 
likely to be immediately extinguished, and the effective temperature is somewhat higher than that given by 
equation \ref{Teff}, owing to the additional luminosity of the companion star circulating within the 
common envelope. 

Equations \ref{rcrit} and \ref{Teff} provide the basis for determining the most likely formation scenario for these Bloatars. 
Table~\ref{bloat_Teff} summarizes the computed effective temperature of the bloated photosphere 
for different combinations of mass/temperature of the primary star and mass/orbital radius of the companion 
planet and shows that a companion Òhot JupiterÓ provides a much more likely match to the observations.

Indeed, if the companion is a star in its own right, the break-up radius is smaller than if it was a hot Jupiter. 
For a companion of mass of 0.21~M$_{\odot}$ orbiting around a primary star of 
$0.4 - 1.0$~M$_{\odot}$, we get $1.35 < R_2/R_{\odot} < 1.56$ and $2800 < T_{\rm eff} < 4700$K. 
At the other extreme, for  a companion of mass of 0.66~M$_{\odot}$ orbiting around a primary 
star of $0.8 - 1.0$~M$_{\odot}$, we get $2.8 < R_2/R_{\odot} < 2.9$ and $3000 < T_{\rm eff} < 3600$K. 
These temperatures are too high to explain the observations. This fact coupled with the long 
inspiral timescale of such massive companions makes it most unlikely that the Bloatars are simply common-envelope binaries.

Even a planet like WASP-18 ( $m_{\rm p} \sim 10M_{\rm J} \sim 10^{-2}$M$_{\odot}$; $r_{\rm p} \sim R_{\rm J} \sim 0.14 R_{\odot}$) 
would likewise provide too high an effective temperature in the common envelope. In the stellar mass range $0.5 - 1.0$ M$_{\odot}$ 
the computed common envelope effective temperature lies in the range $2800 < T_{\rm eff} < 4700$K.

For a standard ``hot Jupiter'' ( $m_{\rm p} \sim 1.0M_{\rm J} \sim 10^{-3}$M$_{\odot}$; $r_{\rm p} \sim R_{\rm J} \sim 0.14 R_{\odot}$) 
the computed common envelope effective temperature lies in the range $1970 < T_{\rm eff} < 3300$K 
over the stellar mass range  $0.5 - 1.0$ M$_{\odot}$. 
However, the tidal heating of the companion planet is likely to bloat the planet up to larger radius, 
causing it to fill its Roche Lobe earlier and at greater radius from the primary star \citep{Leconte10}. 
If the radius of the planet is increased to $r_{\rm p} \sim 0.2R_{\odot}$, then the computed common envelope 
effective temperature lies in the range $1470< T_{\rm eff} < 2490$K over the stellar mass range  $0.5 - 1.0$ M$_{\odot}$.  
This nicely brackets the range of parameters we derive  for our nine stars ($0.5 - 1.0$ M$_{\odot}$ and $1700 < T_{\rm eff} < 2200$K).

Finally we ask: is the envelope formed sufficiently dense to form a ``false photosphere'' 
to the central star? If we adopt the bloated hot Jupiter parameters of the previous paragraph, 
then the common envelope will have a local density in the range $3-6\times10^{-6}$ gm cm$^{-3}$ 
and a column density $\sim 10^{6}$ gm cm$^{-2}$. Thus, although the atmosphere formed is tenuous, 
it has more than enough of a column density to be extremely optically thick to the escape of radiation.

\subsection{Why  have ``Bloatars'' escaped detection so far?} 

The idea of stars ``swallowing planets'' is not new in the literature. For example, \citet{Sie99a} presented simulations of the accretion of massive planets or 
brown dwarfs by asymptotic giant branch stars. In \citet{Sie99b} they also extended this concept to solar-mass stars on the red giant branch. 
Moreover, a few giant and dwarf stars have been suggested to undergo atmospheric expansion and 
swallowing of relatively massive planets in close orbits \citep[see, e.g.,][]{Isr01,Ret03,Li10,Fos10}.

On the other hand,  all these detections involve stellar systems in later evolutionary phases than our 9 Bloatar candidates, which are only a few Myr old. 
Thus, a very important issue we need to address is why Bloatars have not been detected so far in the many photometric and spectroscopic 
surveys of young clusters that are much closer than NGC~3603, i.e. in the solar neighborhood (d$\lesssim$1-2 kpc).  

The probability of detecting a Bloatar is expected to be very small. As we saw in Sect.~\ref{timescale}, 
the ``bloated'' phase would occur at ages comparable with the  first phase of evolution of the planetary system (1-3~Myrs) 
and is expected to last only for about $5 \times 10^4$~yr. 
In NGC~3603, whose rich PMS population and youth provide the circumstances most favorable for the detection of Bloatars, 
they would represent just 1\% of the $\sim$1~Myrs old PMS population. 

These considerations imply that Bloatars could have been found only in nearby star forming regions or clusters younger than a few Myrs. 
The probability of detecting a Bloatar in this restricted sample is much smaller than in a massive cluster like NGC~3603 for several reasons:

\begin{itemize}

\item Nearby 1-3~Myrs old star forming regions and clusters have a typical number of confirmed PMS stars of the order of 10$^2$ rather than 10$^3$ \citep{Rei08a,Rei08b}. 
Thus, if we assume that Bloatars represent just 1\% of the PMS population, only a few of them ($\lesssim$4) are expected to be present in this kind of regions.

\item The spatial extent of nearby star forming regions, like Orion, Serpens, Lupus is of the order of  a few square degrees or more \citep{Rei08a,Rei08b} 
and even more compact young open clusters, like IC~348, typically extend for more than 0.5~deg$^2$ \citep{Mue07}. 
Thus, spectroscopic surveys of the PMS population in these regions are often spatially incomplete and might have easily missed the few Bloatars expected to be present.

\item Photometric surveys for PMS stars and BDs are nowadays spatially complete. However, as we have seen in Sect.~\ref{sel}, broad-band colors of Bloatars are similar to 
those of 0.2-1~M$_{\odot}$ PMS stars. Thus, Bloatars might have been confused with more ordinary PMS stars in such surveys. Only a handful of regions 
have been surveyed combining the use of broad and narrow-band photometry in H$_2$O or CH$_4$ lines \citep[see, e.g.,][]{Naj00,Gor03,Mai04,And06}. 

\end{itemize}


\section{Conclusions \label{conclu}}

We have presented the results of a pioneering narrow-band WFC3/IR survey of NGC~3603 with the aim of 
discovering objects displaying deep water-vapor absorption bands, characteristic of Brown Dwarfs. The main results of our study are as follows:

\begin{itemize}

\item We have developed a method to identify BD candidates on the basis of WFC3/IR narrow-band imaging in water molecular bands.
This photometric method provides effective temperatures for BDs to an accuracy of $\pm 350$K relative to spectroscopic techniques. 
This accuracy is not significantly affected by either stellar surface gravity or uncertainties in the interstellar extinction;

\item The comparison of this calibration relation with expectations from models of stellar/sub-stellar photospheres 
supports a scenario where both dust gravitational settling and scattering/absorption 
operate in the atmosphere of very cool objects;

\item Using this technique, we identify 9 objects having the colors of BDs, strongly clustered towards the luminous core of NGC~3603 but 
far too luminous to be normal BD members of this cluster. We argue that it is unlikely that these objects are either artifacts of our dataset, 
normal field BDs/M-type giants or extra-galactic contaminants and, hence, 
we might have discovered a new class of bloated objects having the effective temperatures of BDs (``Bloatars'');
 
\item We explore the interesting scenario in which these Bloatars would be young stars that have recently tidally ingested a Hot Jupiter, 
the remnants of which is providing a short-lived extended photosphere to the central star. 
We predict that, if this scenario is correct, these stars would show the signature of fast rotation, with the equatorial part of the ``Bloatar'' 
rotating at the orbital velocity of the absorbed planet.

\end{itemize}

\acknowledgments

This paper is based on Early Release Science observations made by 
the WFC3 Scientific Oversight Committee. We are grateful to the 
Director of the Space Telescope Science Institute for awarding 
Director's Discretionary time for this program. 
We thank B. Burningham for providing unpublished spectra of brown dwarfs, X. Pang \& A. Pasquali 
for providing unpublished information on reddening effects in NGC~3603 and the anonymous referee for his careful reading and useful comments/suggestions. 
M. Dopita acknowledges the support of the Australian Research Council (ARC) through Discovery  
projects DP0984657 and DP0664434. 
We also acknowledge extensive use of the SIMBAD database, operated at CDS Strasbourg, the SpeX Prism Spectral Libraries, 
maintained by A. Burgasser, and the Multimission Archive at the Space Telescope Science Institute. 

{\it Facilities:} \facility{HST (WFC3)}.



\end{document}